\documentclass{jfm}

\usepackage{style-ftc}

\shorttitle{Self-similarity in unbounded viscous Marangoni flows}
\shortauthor{F. Temprano-Coleto and H. A. Stone}

\title{On the self-similarity of unbounded viscous Marangoni flows}

\author{Fernando Temprano-Coleto\aff{1,2}\corresp{\email{ftempranocoleto@princeton.edu}} \and H. A. Stone\aff{2}\corresp{\email{hastone@princeton.edu}}}

\affiliation{
\aff{1}Andlinger Center for Energy and the Environment, Princeton University, Princeton, NJ 08544, USA
\aff{2} Department of Mechanical and Aerospace Engineering, Princeton University, Princeton, NJ 08544, USA%
}

\begin{document}

\maketitle

\begin{abstract}
    The Marangoni flow induced by an insoluble surfactant on a fluid-fluid interface is a fundamental problem investigated extensively due to its implications in colloid science, biology, the environment, and industrial applications. Here, we study the limit of a deep liquid subphase with negligible inertia (low Reynolds number, $Re\ll{1}$), where the two-dimensional problem has been shown to be described by the complex Burgers equation. We analyze the problem through a self-similar formulation, providing further insights into its structure and revealing its universal features. Six different similarity solutions are found. One of the solutions includes surfactant diffusion, whereas the other five, which are identified through a phase-plane formalism, hold only in the limit of negligible diffusion (high surface P\'eclet number $Pe_s\gg{1}$). Surfactant `pulses', with a locally higher concentration that spreads outward, lead to two similarity solutions of the first kind with a similarity exponent $\beta=1/2$. On the other hand, distributions that are locally depleted and flow inwards lead to similarity of the second kind, with two different exponents that we obtain exactly using stability arguments. We distinguish between `dimple' solutions, where the surfactant has a quadratic minimum and $\beta=2$, from `hole' solutions, where the concentration profile is flatter than quadratic and $\beta=3/2$. Each of these two cases exhibits two similarity solutions, one earlier than a critical time $t_*$ when the derivative of the concentration is singular, and another one valid after $t_*$. We obtain all six solutions in closed form, and discuss  predictions that can be extracted from these results.
\end{abstract}


\section{Introduction}
When the interface between two fluids is laden with a non-uniform distribution of surfactant, the resulting imbalance of surface tension triggers a Marangoni flow \citep{Scriven1960-ie,De_Gennes2004-hx}. The underlying dynamics are nonlinear, since the surfactant distribution, which sets the surface-driven fluid flow, is itself redistributed by the resulting velocity field through advection. This two-way coupled problem has been the focus of numerous studies, as surface-active molecules are virtually unavoidable in realistic multiphase flow problems appearing both in natural and engineered systems \citep{Manikantan2020-lh}. For example, ambient amounts of surfactant are known to critically alter flows relevant to the environment like the motion of bubbles and drops, through a mechanism first proposed by \cite{Frumkin1947-fm} that has since been studied extensively \citep[][to name a few]{Schechter1963-ne,Wasserman1969-cg,Sadhal1983-fg,Griffith1962-oh,Cuenot1997-mh,Wang1999-kk,Palaparthi2006-ud}. Likewise, the surface of the ocean is affected by surfactants, which alter the dynamics of waves ranging from small capillary ripples \citep{Lucassen1972-nw,Alpers1989-aj} to larger spilling and plunging breakers \citep{Liu2003-vf,Erinin2023-rb}. Marangoni flows also play an important role in biological fluid mechanics, both in physiological transport processes within the lung \citep{Grotberg1995-ph} or the ocular globe \citep{Zhong2019-kd}, and in the motion of colonies of microorganisms that generate biosurfactants \citep{Botte2005-go,Trinschek2018-ev}. In industrially relevant applications, it is well-known that surface-active molecules influence the dip coating of plates and fibers \citep{Park1991-ja,Quere1999-bv}, the drag reduction of superhydrophobic surfaces \citep{Peaudecerf2017-dz,Song2018-hi,Temprano-Coleto2023-hb}, or the stability of foams \citep{Breward2002-si,Cantat2013-se}.

One the most fundamentally important examples of flows induced by surfactants is the so-called `Marangoni spreading' \citep{Matar2009-jt}, where a locally concentrated surfactant spreads unopposed on a clean interface until it reaches a uniform equilibrium concentration. Early quantitative studies examined surfactant spreading on thin films, due to their relevance in pulmonary flows \citep{Ahmad1972-ay,Borgas1988-hd,Gaver1990-wj,Gaver1992-xu}. The pioneering work of \cite{Jensen1992-eq} investigated the spreading of insoluble surfactant from the perspective of \emph{self-similarity}, a powerful theoretical tool to identify universal, scale-free behavior in physical systems \citep{Barenblatt1996-hd}. Several studies of Marangoni flows on thin films based on self-similarity have since followed. For example, \cite{Jensen1993-xa} described the spreading of a soluble surfactant, while \cite{Jensen1994-hf} re-examined the insoluble case, finding additional self-similar solutions for distributions that are not locally concentrated, but depleted of surfactant, which `fill' under the action of Marangoni stresses. Self-similarity was also examined for a deep fluid subphase by \cite{Jensen1995-mu}, considering the limit of dominant fluid inertia (i.e., at high Reynolds number). 

In all the above studies, the problem is simplified by the existence of a confining lengthscale in the fluid subphase, either the thickness of the thin liquid film or the width of the momentum boundary layer. \cite{Thess1995-pm} considered the case of a deep fluid subphase at \emph{low} Reynolds number, where the fluid flow is unconfined, and identified that the resulting problem is \emph{nonlocal}, with the velocity field at any given position depending on the surfactant distribution on the whole interface. Theoretical work in this limit followed \citep{Thess1996-bc,Thess1997-qi} until, recently, \cite{Crowdy2021-vy} showed that the problem is equivalent to a complex version of Burgers equation for a lower-analytic function, effectively providing a \emph{local} reformulation using complex variables. \red{This connection between the nonlocal problem and Burgers equation had also been identified previously \citep{Baker1996-gx,Chae2005-st,De_la_Hoz2008-hp}, albeit in a context unrelated to surfactants or interfacial fluid dynamics. Applied to Marangoni flows, the complex variables formulation has been a key insight to derive new exact solutions \citep{Crowdy2021-vy,Bickel2022-sa} and investigate extensions of the problem \citep{Crowdy2021-ef,Crowdy2023-aj}.}

Even after this simplification \red{for} low-Reynolds-number, deep-subphase Marangoni flow, exact solutions to Burgers equation can be written explicitly only for a selected subset of initial conditions, limiting the generality of the resulting physical insights. In this paper, we analyze the problem from the perspective of self-similarity, which has provided key physical insights not only to Marangoni spreading, but to many other problems like boundary layer theory \citep{Leal2007-zv}, liquid film spreading \citep[e.g.,][]{Huppert1982-nj,Brenner1993-qc,Wu2024-ew}, drop coalescence \citep{Kaneelil2022-kq}, and capillary pinching \citep{Eggers1993-iu,Brenner1996-po,Day1998-tw}. We show that self-similarity not only reveals new universal features about the problem that are independent of the specific \red{initial} conditions, but also gives rise to a beautiful mathematical structure with six different similarity solutions and three different rational exponents, all of which can be obtained in closed form. 

We present the general formulation of the problem in section \ref{sec:problem_formulation}. Section \ref{sec:advect_case} analyzes the case of advection-dominated Marangoni flows, that is, in the limit of infinite surface P\'eclet number $Pe_s^{-1}=0$. In particular, the different possible similarity solutions for this limit are identified through a combination of a phase-plane formalism (\S\,\ref{sec:advect_case_phase_plane}) and stability analysis (\S\,\ref{sec:stability}). In section \ref{sec:spreading_sols}, we consider the case of `spreading', where locally concentrated surfactant induces an outward flow, and derive one solution without diffusion (\S\,\ref{sec:spreading_sols_no_diff}) and one with diffusion (\S\,\ref{sec:spreading_sols_general}). Section \ref{sec:filling_sols} analyzes locally depleted surfactant distributions, which induce a `filling' flow inwards. Depending on the initial conditions, we distinguish that the filling dynamics converge to either `dimple' (\S\,\ref{sec:filling_sols_dimple}) or `hole' (\S\,\ref{sec:filling_sols_hole}) solutions. For either case, we derive one similarity solution that holds prior to a reference time $t_*$ where the solution has a singularity, and another similarity solution valid after $t_*$. We discuss these results and draw conclusions in section \ref{sec:conclusions}.

\section{Problem formulation} \label{sec:problem_formulation}
\subsection{Governing equations}
\label{sec:problem_formulation_gov-eqns}
We consider the dynamics of an insoluble surfactant evolving on the free surface of a layer of incompressible, Newtonian fluid of density $\rho$ and dynamic viscosity $\mu$. Our focus is the limit of small Reynolds ($Re$) and capillary ($Ca$) numbers given by
\begin{equation}
    Re = \dfrac{\rho{u_c}{l_c}}{\mu}\ll{1}, \qquad\qquad Ca = \dfrac{\mu{u_c}}{\gamma_0}\ll{1},
\end{equation}
where $\gamma_0$ is the surface tension of the clean (surfactant-free) interface, and $l_c$ and $u_c$ are the characteristic length and velocity scales of the problem, respectively. In this asymptotic limit, surface tension dominates over viscous stresses, keeping the interface flat. In addition, fluid inertia is negligible and the velocity field $\bs{u}(\bs{x},t)$, which depends on both time $t$ and position $\bs{x}$, is well described by the continuity and Stokes equations
\begin{equation}
    \divv\bs{u}=0, \qquad\qquad \divv\tenss{\sigma}={\bf 0},
    \label{eq:stokes}
\end{equation}
where $\tenss{\sigma}$ is the second-order stress tensor,  $\tenss{\sigma}=-p\,\tens{I} + \mu\bigl(\gradd\bs{u}+\trans{\left(\gradd\bs{u}\right)}\bigr)$, $p$ the mechanical pressure, and $\tens{I}$ the identity tensor.
\begin{figure}
    \centering 
    \subfloat{\includegraphics[]{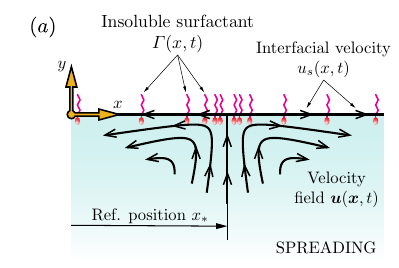}\label{fig:problem_setup_spreading}}
    \subfloat{\includegraphics[]{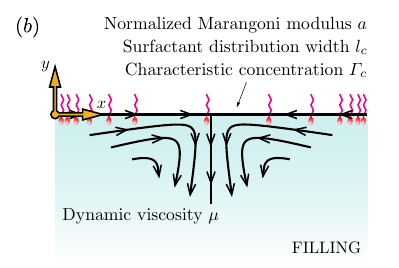}\label{fig:problem_setup_filling}}
    \caption{A non-homogeneous distribution of insoluble surfactant, with concentration $\Gamma(x,t)$, induces a velocity field $\bs{u}(\bs{x},t)$ within its fluid subphase through interfacial Marangoni stresses. The surfactant is itself advected by the resulting interfacial velocity $u_s(x,t)=\left.\bs{u}\bs{\cdot}\bs{e}_x\right|_{y=0}$, leading to a two-way coupled problem. $(a)$ For a localized \emph{pulse} of surfactant, the Marangoni flow results in outward `spreading'. $(b)$ When the surfactant distribution is instead depleted at its center (a \emph{hole} or a \emph{dimple}), the result is an inward `filling' flow. All the dimensional parameters of the model considered here are highlighted in panel $(b)$.}
    \label{fig:problem_setup}
\end{figure}

For sufficiently elongated surfactant distributions (e.g., a `strip' of surfactant) and a sufficiently deep fluid subphase, the problem can be reduced to the unbounded, two-dimensional scenario displayed in figure \ref{fig:problem_setup}. We use a coordinate system where $x$ spans the interface and $y$ points away from the fluid subphase, with $\bs{e}_x$ and $\bs{e}_y$ the unit vectors in the $x$ and $y$ directions, respectively. Velocity components are denoted $u$ and $v$, with $\bs{u}=u\bs{e}_x+v\bs{e}_y$. The domain is considered to be semi-infinite, defined in $x\in(-\infty,\infty)$ and $y\in(-\infty,0\,]$, and the time evolution of the surfactant concentration $\Gamma(x,t)$ along the interface is given by
\begin{equation}
    \pd{\Gamma}{t}{} + \pd{(u_s\Gamma)}{x}{} = D_s\pd{\Gamma}{x}{2}, 
    \label{eq:surfactant_pde}
\end{equation}
where $D_s$ is the surface diffusivity of the surfactant and $u_s = \left.\bs{u}\bs{\cdot}\bs{e}_x\right|_{y=0}$ is the interfacial velocity. The boundary conditions at the interface 
\begin{subequations}
    \begin{align}        
    \left.\bs{e}_x\bs{\cdot}\tenss{\sigma}\bs{\cdot}\bs{e}_y\right|_{y=0} &=  \red{-}a\pd{\Gamma}{x}{}, \label{eq:BCs:Marangoni}\\
        v(y=0) &= 0, \label{eq:BCs:no_penetr}
    \end{align}
    \label{eq:BCs}
\end{subequations}
are the Marangoni condition \eqref{eq:BCs:Marangoni} linking viscous stresses to the gradient of surfactant, and the no-penetration kinematic condition \eqref{eq:BCs:no_penetr}. Here, the parameter 
\begin{equation}
    \red{a =\left|\td{\gamma}{\Gamma}{}\right|}
\end{equation}
is a normalized Marangoni modulus, indicating the variation of surface tension $\gamma$ with respect to surfactant concentration. We regard $D_s$ and $a$ as constants, although they are in general dependent on $\Gamma$ through equations of state $D_s(\Gamma)$ and $\gamma(\Gamma)$, e.g., see \cite{Manikantan2020-lh}.

The governing equations \eqref{eq:stokes}-\eqref{eq:surfactant_pde} and boundary conditions \eqref{eq:BCs} are supplemented with an initial condition for the surfactant distribution
\begin{equation}
    \Gamma(x,t=0) = \Gamma_0(x).
    \label{eq:IC}
\end{equation}
The profile $\Gamma_0(x)$ introduces the characteristic scale $\Gamma_c$, which we will take as the maximum concentration $\Gamma_c = \max_x{[\Gamma_0(x)]}$. Furthermore, the typical width of $\Gamma_0(x)$  sets the lengthscale $l_c$ of the problem.

From these constants, dimensional analysis of the Marangoni boundary condition \eqref{eq:BCs:Marangoni} leads to a natural scale for the velocity magnitude
\begin{equation} 
    \left.\bs{e}_x\bs{\cdot}\tenss{\sigma}\bs{\cdot}\bs{e}_y\right|_{y=0} = \left.\mu\pd{u}{y}{}\right|_{y=0} = \red{-}a\pd{\Gamma}{x}{} \implies u_c \approx \dfrac{a\Gamma_c}{\mu}.
\end{equation}
For the assumptions of $Re\ll{1}$ and $Ca\ll{1}$ to hold, the characteristic concentration $\Gamma_c$ and width $l_c$ of the surfactant distribution must both be sufficiently small to ensure that
\begin{equation}
    l_c\Gamma_c \ll \dfrac{\mu^2}{\rho{a}}, \qquad\qquad \Gamma_c \ll \dfrac{\gamma_0}{a},
\end{equation}
providing practical estimates to determine if, for a given set of physicochemical properties $\mu$, $\rho$, $a$, and $\gamma_0$, a known surfactant distribution will lead to Marangoni flow in the asymptotic limit considered in this study.

The full problem, as defined by \eqref{eq:stokes}-\eqref{eq:BCs} and \eqref{eq:IC}, is nonlinear and involves the two-dimensional vector field $\bs{u}$. \cite{Thess1995-pm} recognized that it was possible to obtain a one-dimensional formulation, using the Fourier transform of \eqref{eq:stokes} to obtain $u_s$ as a function of $\Gamma$. Here, we show that the same simplification can be achieved through the boundary integral representation of Stokes flow. Indeed, the velocity field given by \eqref{eq:stokes} at any position $\bs{x}$ along the interface \citep[see][]{Pozrikidis1992-gr} can be expressed as
\begin{equation}
    \red{\bs{u}(\bs{x},t) = \dfrac{1}{2\pi}\left[\int_\mathcal{I}\tens{G}(\bs{x}-\bs{x}')\bs{\cdot}\tenss{\sigma}(\bs{x}',t)\bs{\cdot}\bs{n}\dd{l_{\bs{x}'}}+\dashint_\mathcal{I}\bs{u}(\bs{x}',t)\bs{\cdot}\tens{T}(\bs{x}-\bs{x}')\bs{\cdot}\bs{n}\dd{l_{\bs{x}'}}\right]},
    \label{eq:boundary_integral}
\end{equation}
where the dash denotes the Cauchy principal value of the integral, \red{$\mathcal{I}$ denotes the interface}, $\bs{n}$ the unit outward normal vector, and the tensors in the integrands are defined as
\begin{subequations}
    \begin{align}
        \tens{G}(\bs{x}) &= -\dfrac{1}{\mu}\left(\ln|\bs{x}|\,\tens{I}-\dfrac{\bs{x}\bs{x}}{|\bs{x}|^2}\right),\\
        \tens{T}(\bs{x}) &= -4\,\dfrac{\bs{x}\bs{x}\bs{x}}{|\bs{x}|^4}.
    \end{align}
    \label{eq:tensors}
\end{subequations}
Since the interface remains flat, the outward normal vector simplifies as $\bs{n}=\bs{e}_y$, while $\bs{x}$ and $\bs{x}'$ are co-linear and parallel to $\bs{e}_x$. Therefore, we have $(\bs{x}-\bs{x}')\bs{\cdot}\bs{n}=0$ since the vectors are orthogonal, and
\begin{equation}
    \tens{T}(\bs{x}-\bs{x}')\bs{\cdot}\bs{n} = -4\,\dfrac{(\bs{x}-\bs{x}')(\bs{x}-\bs{x}')(\bs{x}-\bs{x}')}{|\bs{x}-\bs{x}'|^4}\bs{\cdot}\bs{n} = \tens{0},
\end{equation}
eliminating the \red{second} integral (the `\red{double-layer} potential') in \eqref{eq:boundary_integral}. Taking $(\bs{x}-\bs{x}')=(x-x')\bs{e}_x$ and noting the integrals along the interface are simply along $x'$, the interfacial velocity can then be expressed as
\begin{equation}
    u_s(x,t) = \left.\bs{e}_x\bs{\cdot}\bs{u}\right|_{y=0} = \dfrac{1}{2\pi\mu}\int_{-\infty}^{\infty}\Bigl(1-\ln|x-x'|\Bigr)\left.\bs{e}_x\bs{\cdot}\tenss{\sigma}(x',t)\bs{\cdot}\bs{e}_y\right|_{y=0}\dd{x'},
\end{equation}
which, upon substitution of the Marangoni boundary condition \eqref{eq:BCs:Marangoni} and integration by parts, becomes
\begin{align}
    u_s(x,t) = \dfrac{a}{2\pi\mu}&\left\{\Bigl[\bigl(\ln|x-x'|-1\bigr)\,\Gamma(x',t)\Bigr]_{x'\to-\infty}^{x'\to\infty} +\dashint_{-\infty}^{\infty}\dfrac{\Gamma(x',t)}{x-x'}\dd{x'}\right\}.
    \label{eq:u_s_long}
\end{align}
\red{The first term in \eqref{eq:u_s_long} vanishes for any surfactant profile decaying as a power law (or faster) in the far field. Remarkably, this term also cancels out for profiles $\Gamma(x,t)$ that do not decay as $|x|\to\infty$, as long as their far-field values are finite and symmetric, such that $0<\lim_{x\to\infty}\Gamma(x,t) = \lim_{x\to-\infty}\Gamma(x,t)<\infty$. We therefore restrict this study to these two possible far-field behaviors, excluding `step-like' profiles with asymmetric far-field concentrations}. The above leads to the closure relationship 
\begin{align}
    u_s(x,t) = \dfrac{a}{2\pi\mu}\dashint_{-\infty}^{\infty}\dfrac{\Gamma(x',t)}{x-x'}\dd{x'} = \dfrac{a}{2\mu}\mathcal{H}[\Gamma],
    \label{eq:closure_u_s}
\end{align}
first derived by \cite{Thess1995-pm}, where the operator $\hil{\,\,\,}$ denotes the Hilbert transform of a function \citep[for details, see][]{King2009-hr,King2009-bq}. This closure relationship results in a one-dimensional problem, only requiring the solution of \eqref{eq:surfactant_pde} alongside  condition \eqref{eq:closure_u_s}. The resulting formulation is, however, \emph{nonlocal}, as the interfacial velocity $u_s$ at any given point depends upon the distribution of $\Gamma$ along the whole real line.

We proceed to nondimensionalize equations \eqref{eq:surfactant_pde}, \eqref{eq:closure_u_s}, and the initial condition \eqref{eq:IC} using the scales of the problem discussed above. To that end, we apply the rescalings
\begin{equation}
    x \to l_c{x}, \qquad t \to \left(\dfrac{2\mu{l_c}}{a\Gamma_c}\right)t, \qquad u \to \left(\dfrac{a\Gamma_c}{2\mu}\right)u, \qquad \Gamma \to \Gamma_c\Gamma, \qquad \Gamma_0 \to \Gamma_c\Gamma_0,
\end{equation}
which leads to a dimensionless problem given by
\begin{subequations}
    \begin{align}
        \pd{\Gamma}{t}{} + \pd{(\hil{\Gamma}\Gamma)}{x}{} &= \dfrac{1}{Pe_s}\pd{\Gamma}{x}{2}, \label{eq:nondim_Gamma}\\
        \Gamma(x,t=0) &= \Gamma_0(x), \label{eq:nondim_eqns_IC}
    \end{align}
    \label{eq:nondim_eqns}
\end{subequations}
and where the surface P\'eclet number is defined as
\begin{align} \label{eq:peclet}
    Pe_s \defeq \dfrac{a\Gamma_cl_c}{2\mu{D_s}}.
\end{align}

\red{Alternative formulations of the nonlocal problem \eqref{eq:nondim_eqns} have been studied in the mathematical literature \citep{Baker1996-gx,Morlet1998-rn,Chae2005-st,De_la_Hoz2008-hp,Eggers2020-wo} as a model for finite-time blowup. However, that body of work was not concerned with the description surfactant dynamics, which introduces key differences that we highlight in section \ref{sec:problem_formulation_others} below. In the context of Marangoni flows, \cite{Crowdy2021-vy} recently showed, through a complex-variable formulation of two-dimensional Stokes flow, that a dependent variable $\psi=u_s+\ii\Gamma = \hil{\Gamma} + \ii\Gamma$ satisfies}
\begin{subequations}
    \begin{align}
        \pd{\psi}{t}{} + \psi\pd{\psi}{x}{} &= \dfrac{1}{Pe_s}\pd{\psi}{x}{2}, \label{eq:Burgers_pde}\\
        \psi(x,t=0) &= \psi_0(x) = \red{\hil{\Gamma_0(x)}+\ii\Gamma_0(x)}, \label{eq:Burgers_IC}
\end{align}
    \label{eq:Burgers}
\end{subequations}
\red{where $\psi(z,t)$ must be a \red{lower}-analytic complex function \citep[named $h(z,t)$ in the notation of][]{Crowdy2021-vy} of the variable $z=x+\ii{y}$. This problem reduction can also be realized by adding \eqref{eq:nondim_Gamma} to its Hilbert transform, and then recognizing that $\hil{\partial_x\Gamma}=\partial_x\hil{\Gamma}$ and $\hil{\Gamma\hil{\Gamma}}=((\hil{\Gamma})^2-\Gamma^2)/2$. It is worth remarking that \emph{subtracting} \eqref{eq:nondim_Gamma} from its Hilbert transform leads to the same Burgers equation \eqref{eq:Burgers_pde} but with a complex conjugate dependent variable $\psi=u_s-\ii\Gamma$, which is an equivalent notation followed by \cite{Bickel2022-sa}. In such an alternative notation the complex function $\psi(z,t)$ is instead the upper-analytic Schwarz conjugate of the one used here.} 

The limit of negligible diffusion given by $Pe_s\gg{1}$ can  be approximated at leading order by taking $Pe_s^{-1}=0$, which yields 
\begin{subequations}
    \begin{align}
        \pd{\psi}{t}{} + \psi\pd{\psi}{x}{} &= 0, \label{eq:Hopf_pde}\\
        \psi(x,t=0) &= \psi_0(x) = \red{\hil{\Gamma_0(x)}+\ii\Gamma_0(x)}. \label{eq:Hopf_IC}
\end{align}
    \label{eq:Hopf}
\end{subequations}
The problems given by the Burgers equation \eqref{eq:Burgers_pde} and the inviscid Burgers equation \eqref{eq:Hopf_pde} (also known as the Hopf equation) are now \emph{local}, and admit exact solutions via either the Cole-Hopf transformation for \eqref{eq:Burgers} or the method of characteristics for \eqref{eq:Hopf}, as shown in \cite{Crowdy2021-vy} and \cite{Bickel2022-sa}. While some of these solutions have been shown to exhibit self-similar behavior \citep{Thess1996-bc,Bickel2022-sa}, a systematic analysis of the problem from the perspective of self-similarity has not yet been performed, and is the goal of this paper.

\subsection{Self-similar formulation} \label{sec:problem_formulation_self-sim}
We adopt the following self-similarity ansatzes:
\begin{subequations} \label{eq:ansatz}
\begin{align}
    \eta &= \sgn{t-t_*}\dfrac{x-x_*}{A\,|t-t_*|^\beta}, \label{eq:ansatz_eta}\\
    \psi(x,t) &= \red{B}\,|t-t_*|^\alpha\, f(\eta), \label{eq:ansatz_psi}
\end{align}
\end{subequations}
with $\eta$ the real similarity variable. \red{T}he similarity function $f(\eta)$, which takes complex values, \red{is decomposed} as $\red{f = \U + \ii\C}$, with $U(\eta)$ and $C(\eta)$ real. The interfacial velocity and surfactant concentration can then be recovered as
\begin{subequations} \label{eq:us_Gamma_f}
\begin{align}
    u_s(x,t) &= \red{B}\,|t-t_*|^{\alpha}\,\U(\eta), \label{eq:us_f}\\
    \Gamma(x,t) &= \red{B}\,|t-t_*|^{\alpha}\, \C(\eta).\label{eq:Gamma_f}
\end{align}
\end{subequations}

\red{We introduce the positive real constants $A$ and $B$ for convenience, and fix their values to simplify the final form of the similarity solutions $f(\eta)$, as illustrated below}. The real constants $x_*$ and $t_*$ are a reference position and time, respectively. Including the factor $\sgn{t-t_*}$ in the definition \eqref{eq:ansatz_eta} is equivalent to choosing
\begin{subequations}
    \begin{align}
        \eta &=\dfrac{x-x_*}{A(t-t_*)^\beta} \label{eq:eta_fwd}
\intertext{for solutions that evolve forward in time $t>t_*$, and to choosing}
        \eta &=\dfrac{x_*-x}{A(t_*-t)^\beta} \label{eq:eta_bwd}
\end{align}
\end{subequations}
for solutions that evolve backward in time $t<t_*$. We adopt the more intuitive forward-time description when describing `spreading' (as in figure \ref{fig:problem_setup_spreading}) solutions of \eqref{eq:Burgers} and \eqref{eq:Hopf}. These solutions become self similar at long times $t\gg{1}$, so in that case we take $t_*=0$. However, `filling' self-similar solutions representing inward flow (as in figure \ref{fig:problem_setup_filling}) are often only valid sufficiently close to a reference time $t=t_*>0$ \red{at which} the solution has a singularity, requiring either the backward-time \citep[e.g.,][]{Eggers2008-lk} or the forward-time \citep[e.g.,][]{Zheng2018-bu} description to analyze their behavior immediately prior or subsequent to $t_*$, respectively.

Using the self-similarity ansatzes \eqref{eq:ansatz} in Burgers equation \eqref{eq:Burgers_pde} leads to
\begin{align} \label{eq:ode_f_exponents}
    \alpha{f}-\beta\eta\td{f}{\eta}{} + \red{\dfrac{B}{A}}|t-t_*|^{\alpha-\beta+1}f\td{f}{\eta}{} = \dfrac{\sgn{t-t_*}}{A^2Pe_s}|t-t_*|^{1-2\beta}\td{f}{\eta}{2}.
\end{align}
Solutions are self-similar when the above ODE is solely dependent on $\eta$, and not on $t$ or $x$ separately, which requires either one of the following two scenarios:
\begin{enumerate}
    \item For the general case of a finite $Pe_s^{-1}>0$, the only possible choice of exponents is $\alpha=-1/2$, $\beta=1/2$. \red{Seeking to eliminate parameters from equation \eqref{eq:ode_f_exponents}, we fix $A=B=\sqrt{2/Pe_s}$, focusing on forward-time solutions with $\sgn{t-t_*}=1$ that lead to}
    \begin{equation}
        \td{}{\eta}{}\left[\td{f}{\eta}{}+\eta{f}-f^2\right]=0.
        \label{eq:ode_f_full}
    \end{equation}
    Equation \eqref{eq:ode_f_full} has one spreading (as in figure \ref{fig:problem_setup_spreading}) self-similar solution of the first kind, which was identified by \cite{Bickel2022-sa} and which we outline in section \ref{sec:spreading_sols_general}.
    \item In the advection-dominated limit given by $Pe_s^{-1}=0$, self-similarity only requires $\alpha=\beta-1$\red{. In this case, we fix $B=A$, keeping} $\beta$ and $A$ \red{as} free parameters. This leads to
    \begin{align}
        (f-\beta\eta)\td{f}{\eta}{} = (1-\beta){f}.
        \label{eq:ode_f_no_diff}
    \end{align}
    We obtain the same similarity equation \eqref{eq:ode_f_no_diff} independently of the choice of the forward-time or backward-time definition of $\eta$, due to the invariance of the inviscid Burgers equation \eqref{eq:Hopf_pde} with respect to a reversal of time $t\to{-t}$ and space $x\to{-x}$. In this advection-dominated case, multiple solutions can potentially arise, depending on the specific value of $\beta$. Using a phase-plane formalism and stability analysis, in section \ref{sec:advect_case} we identify five possible similarity solutions of \eqref{eq:ode_f_no_diff}. We re-discover the spreading self-similar solution of the first kind first identified by \cite{Thess1996-bc}, which we detail in section \ref{sec:spreading_sols_no_diff}. We also find four possible `filling' (as in figure \ref{fig:problem_setup_filling}) solutions of the second kind with different power-law exponents $\beta$, which we describe in section \ref{sec:filling_sols}.
\end{enumerate}

For either of the two similarity equations \eqref{eq:ode_f_full} and \eqref{eq:ode_f_no_diff} above, solutions $f(\eta)$ must have a physically correct parity. \red{Introducing the similarity ansatzes \eqref{eq:ansatz} into the closure relationship $u_s(x,t)=\hil{\Gamma(x,t)}$ leads to an analogous relation $U(\eta)=\hil{C(\eta)}$ for the similarity solutions. Since the Hilbert transform reverses parity \citep{King2009-hr}, we can conclude that the only admissible parities are either $U$ odd and $C$ even, or $U$ even and $C$ odd.} However, an odd function $C(\eta)$ would imply unphysical negative values of the concentration $\Gamma(x,t)$. Accordingly, we only consider similarity solutions with $U(\eta)$ odd and $C(\eta)$ even \red{or, equivalently, $f(-\eta)=-\overline{f(\eta)}$ with the overbar indicating complex conjugation}. Note that this parity requirement does not necessarily apply to the physical solutions $\Gamma(x,t)$ and $u_s(x,t)$ which, as we show in sections \ref{sec:spreading_sols} and \ref{sec:filling_sols}, can be asymmetric and only attain symmetry as they converge to a self-similar solution.

In addition, similarity solutions $f(\eta)$ must satisfy a specific far-field boundary condition \citep{Eggers2008-lk} \red{such} that the function $\psi(x,t)$ is independent of time in the far field $|x|\to\infty$. From the similarity ansatz \eqref{eq:ansatz_psi} and from the fact that $\alpha=\beta-1$, it is clear that such a far-field behavior requires \red{$f(\eta) =O(|t-t_*|^{1-\beta})$ as $|\eta|\to\infty$}. Given the definition of $\eta$ in \eqref{eq:ansatz_eta}, the only possibility to satisfy \red{this} condition is
\begin{align} \label{eq:far_field}
    f(\eta) \sim \red{k_{\pm\infty}} |\eta|^\frac{\beta-1}{\beta} \quad\text{as }|\eta|\to\infty.
\end{align}
We use the notation $\red{k_{\pm\infty}}$ to \red{denote generic} far-field constants \red{that} differ between $\red{k_{\infty}}$ as $\eta\to\infty$ and $\red{k_{-\infty}}$ as $\eta\to-\infty$ since, by symmetry, $\red{k_{-\infty}=-\bar{k}_{\infty}}$. Equation \eqref{eq:far_field} is referred to as a `quasi-stationary' far-field condition. If the similarity solution $f(\eta)$ is globally valid in space, then the condition is equivalent to a far-field behavior of $\psi$ that is constant in time as $|x|\to\infty$. If, on the other hand, $f(\eta)$ is only valid locally (as is often the case with similarity of the second kind), the condition implies that $f(\eta)$ must match with the `outer' non-self-similar part of $\psi$, which evolves on a slower time scale. 

\subsection{\red{A note on alternative formulations}}
\label{sec:problem_formulation_others}
\red{The problem given by \eqref{eq:nondim_eqns} and \eqref{eq:Burgers}, as well as its variants with $Pe_s^{-1}=0$, appear in the literature with slightly different formulations that are nonetheless equivalent, which we summarize in table \ref{tab:formulations}. It is easy to check that, \emph{once transformed to our formulation}, the finite-time blowup described by the mathematical literature \citep[][]{Chae2005-st,Baker1996-gx,Morlet1998-rn,De_la_Hoz2008-hp,Eggers2020-wo} occurs for \emph{negative} concentration $\Gamma<0$. This is unphysical if $\Gamma$ represents a surfactant concentration, but it is not problematic in the above body of work, where the nonlocal problem \eqref{eq:nondim_eqns} has an unrelated physical motivation. Those studies describe singularities that are persistent in time and occur at points with $\Gamma<0$, thus qualitatively different from those found here (section \ref{sec:filling_sols}), which happen at points of $\Gamma=0$ within a non-negative profile $\Gamma(x)\geq{0}$, and disappear at finite time. For that reason, the self-similar analysis by \cite{De_la_Hoz2008-hp} and \cite{Eggers2020-wo} is \emph{linearized} around the nonzero value of $\Gamma$ at which singularities occur, yielding different similarity equations and exponents.}
\begin{table}
\begin{center}
\def~{\hphantom{0}}
\begin{tabular}{c c c c}
     \multirow{3}{*}{Nonlocal form} & \multirow{3}{*}{Local form} & Conversion & \multirow{3}{*}{Examples}\\
     & & to \eqref{eq:nondim_eqns} & \\
     & & and \eqref{eq:Burgers} & \\
     \hline
     \multirow{3}{*}{\shortstack[c]{$\Gamma_t + (\hil{\Gamma}\Gamma)_x = Pe_s^{-1}\Gamma_{xx}$\\ $\Gamma(x,0)=\Gamma_0(x)$}} & \multirow{3}{*}{\shortstack[c]{$\psi_t + \psi\psi_x = Pe_s^{-1}\psi_{xx}$\\ $\psi(x,0)=\hil{\Gamma_0}+\ii\Gamma_0$}} & \multirow{3}{*}{None} & This work,\\
     & & & \cite{Chae2005-st}, \\
     & & & \cite{Crowdy2021-vy}\\
     \hline
     \multirow{3}{*}{\shortstack[c]{$\Gamma_t + (\hilalt{\Gamma}\Gamma)_x = Pe_s^{-1}\Gamma_{xx}$\\ $\Gamma(x,0)=\Gamma_0(x)$}} & \multirow{3}{*}{\shortstack[c]{$\psi_t + \psi\psi_x = Pe_s^{-1}\psi_{xx}$\\ $\psi(x,0)=\hilalt{\Gamma_0}-\ii\Gamma_0$}} & $\widetilde{\mathcal{H}}\to-\mathcal{H}$ & \multirow{3}{*}{\cite{Baker1996-gx}} \\
     & & $\Gamma\to-\Gamma$ & \\
     & & $\Gamma_0\to-\Gamma_0$& \\
     \hline
     $\Gamma_t - \frac{1}{2}(\hilalt{\Gamma}\Gamma)_x = 0$ & \multirow{2}{*}{N/A} & $\widetilde{\mathcal{H}}\to-\mathcal{H}$ & \multirow{2}{*}{\cite{Thess1996-bc}}\\
     $\Gamma(x,0)=\Gamma_0(x)$ & & $t\to2t$ & \\
     \hline
     $\Gamma_t - (\hilalt{\Gamma}\Gamma)_x = Pe_s^{-1}\Gamma_{xx}$ & \multirow{2}{*}{N/A} & \multirow{2}{*}{$\widetilde{\mathcal{H}}\to-\mathcal{H}$} & \multirow{2}{*}{\cite{Morlet1998-rn}}\\
     $\Gamma(x,0)=\Gamma_0(x)$ & & & \\
     \hline
     \multirow{3}{*}{\shortstack[c]{$\Gamma_t - (\hil{\Gamma}\Gamma)_x = 0$\\$\Gamma(x,0)=\Gamma_0(x)$}} & \multirow{3}{*}{\shortstack[c]{$\psi_t-\psi\psi_x=0$\\$\psi(x,0)=\hil{\Gamma_0}+\ii\Gamma_0$}} & $\Gamma\to-\Gamma$ & \multirow{3}{*}{\shortstack[c]{\cite{De_la_Hoz2008-hp},\\\cite{Eggers2020-wo}}} \\
     & & $\Gamma_0\to-\Gamma_0$ & \\
     & & $\psi\to-\psi$ & \\
     \hline
     $\Gamma_t + (\hil{\Gamma}\Gamma)_x = Pe_s^{-1}\Gamma_{xx}$ & $\psi_t + \psi\psi_x = Pe_s^{-1}\psi_{xx}$ & \multirow{2}{*}{$\psi\to\overline{\psi}$} & \multirow{2}{*}{\cite{Bickel2022-sa}}\\
     $\Gamma(x,0)=\Gamma_0(x)$ & $\psi(x,0)=\hil{\Gamma_0}-\ii\Gamma_0$ & & \\
     \hline
\end{tabular}
\caption{\red{Alternative formulations of the problem found in the literature, alongside the transformations required to convert them to \eqref{eq:nondim_eqns}, \eqref{eq:Burgers}, and their variants with $Pe_s^{-1}=0$. Here, $\hilalt{\Gamma} = \pi^{-1}\dashint_{-\infty}^\infty(x'-x)^{-1}\Gamma(x',t)\,\dd{x'}$ is an alternative definition of the Hilbert transform, such that $\hilalt{\Gamma}=-\hil{\Gamma}$. Subindices indicate partial derivatives.}}
\label{tab:formulations}
\end{center}
\end{table}

\section{Analysis of the advection-dominated case} \label{sec:advect_case}
In the advection-dominated case with $Pe_s^{-1}=0$, the complexity of the similarity ODE \eqref{eq:ode_f_no_diff} can be reduced by noting that it is scale-invariant, since the transformations $f \to \lambda{f}$ and $\eta \to \lambda\eta$ (with $\lambda$ real and nonzero) leave the equation unchanged. The ratio $f/\eta\to \lambda{f}/(\lambda{\eta}) = f/\eta$ also remains invariant under these rescalings, suggesting a change of dependent variable $g(\eta) \defeq f(\eta)/\eta$ that turns equation \eqref{eq:ode_f_no_diff} into
\begin{align}
     \td{g}{\ln|\eta|}{} = \dfrac{g(1-g)}{(g-\beta)}{}.
    \label{eq:ode_g}
\end{align}
Equation \eqref{eq:ode_g} is a \emph{separable} first-order ODE, so it can be integrated directly to obtain
\begin{equation}
    \red{
    \left(1-\dfrac{f}{\eta}\right)\left(\dfrac{f}{\eta}\right)^{-\frac{\beta}{\beta-1}} = k_\pm|\eta|^\frac{1}{\beta-1},
    }
    \label{eq:implicit_sol_f}
\end{equation}
\red{where exponentiation of complex numbers is understood in a principal value sense. We again write $k_\pm$ to highlight that, for complex solutions, the complex integration constant can in principle take different values $k_+$ for $\eta>0$, and $k_-$ for $\eta<0$. In fact, introducing the transformations $\eta\to-\eta$ and $f\to-\bar{f}$ in \eqref{eq:implicit_sol_f}, we find that
\begin{equation}
    \red{
    k_-=\bar{k}_+
    }
    \label{eq:k_condition}
\end{equation}
for solutions to have the required symmetry $f(-\eta)=-\overline{f(\eta)}$.}

\red{Equation} \eqref{eq:implicit_sol_f} is still an implicit relation, providing little insight \red{about solutions} for arbitrary real values of $\beta$. It is therefore not straightforward, at least from \eqref{eq:implicit_sol_f} alone, to determine the particular subset of physically realistic similarity solutions.

\subsection{The phase plane} \label{sec:advect_case_phase_plane}
Since equation \eqref{eq:ode_g} is also \emph{autonomous}, its solutions can be represented in a phase plane with state variables $\re{g}=U/\eta$ and $\red{\im{g}=C/\eta}$, following \cite{Gratton1990-uc}. This formalism allows systematic identification of all possible similarity solutions of \eqref{eq:ode_f_no_diff} as distinct trajectories in the phase plane\red{, with the beginning of each trajectory representing the origin $\eta=0$ and its end point indicating the far field $|\eta|\to\infty$}. We construct the phase plane by first finding the fixed points of equation \eqref{eq:ode_g}, seeding initial conditions closely around each of them, and then numerically integrating forward or backward in $\ln|\eta|$ depending on if the particular seed is along a stable or unstable direction. A detailed account of the integration procedure and the calculation of the fixed points is provided in appendix \ref{sec:apdx_phase_plane}. We only consider exponents $\beta>0$, excluding also $\beta=1$ since it leads to a linear problem in $\eqref{eq:ode_g}$ with constant solutions for $f(\eta)$. 
\begin{figure}
    \centering 
    \subfloat{\includegraphics[]{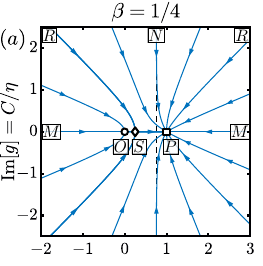}\label{fig:phase_plane_a}}
    \subfloat{\includegraphics[]{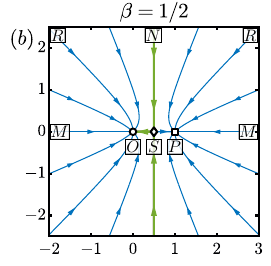}\label{fig:phase_plane_b}}
    \subfloat{\includegraphics[]{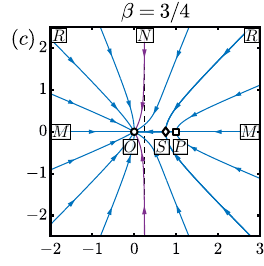}\label{fig:phase_plane_c}}\\[-10pt]
    \subfloat{\includegraphics[]{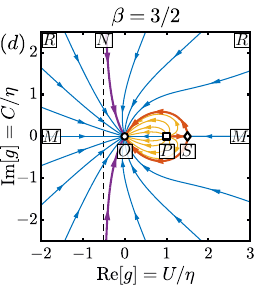}\label{fig:phase_plane_d}}
    \subfloat{\includegraphics[]{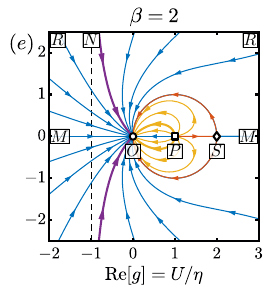}\label{fig:phase_plane_e}}
    \subfloat{\includegraphics[]{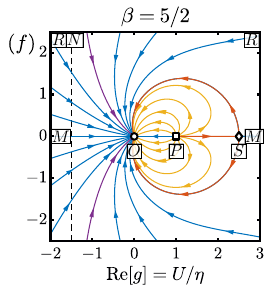}\label{fig:phase_plane_f}}
    \caption{Phase portraits of equation \eqref{eq:ode_g}, for six different values of $\beta>0$, $\beta\neq{1}$. Any given two trajectories that are symmetric with respect to the horizontal axis represent one possible similarity solution $f(\eta)=\eta\,{g(\eta)}$, with the origin of the trajectory denoting $\eta=0$ and the endpoint denoting $|\eta|\to\infty$. The three fixed points $O=(0,0)$ (stable node), $P=(1,0)$ (node), and $S=(\beta,0)$ (saddle) have horizontal and vertical eigendirections for all $\beta>0$. Points $M$, $N$, $R$ are the fixed points of the ODE satisfied by the reciprocals of the solution (see appendix \ref{sec:apdx_phase_plane}). Only green, purple, orange, and yellow trajectories represent similarity solutions that are physically relevant, as illustrated in table \ref{tab:trajectories}. Stability criteria (section \ref{sec:stability}) select the only five solutions that can be obtained in practice, which are highlighted with a wider streak. The dashed vertical line corresponds to $U/\eta=1-\beta$.}
    \label{fig:phase_plane}
\end{figure}

Phase portraits of the system are shown in figure \ref{fig:phase_plane}, for a set of six representative values of $\beta$. The phase plane has a remarkably simple structure, being symmetric with respect to the horizontal axis $C/\eta=0$ due to the invariance of  \eqref{eq:ode_g} to $g\to\bar{g}$. \red{Note that the symmetry $f(-\eta)=-\overline{f(\eta)}$ required of the similarity solution leads to $g(-\eta)=\overline{g(\eta)}$, given that $f(\eta)=\eta\,{g}(\eta)$. In the phase plane, this means that a full solution is represented by a \emph{combination} of a trajectory in the upper half-plane (given by $g$) and and its reflection in the lower half-plane (given by $\overline{g}$). The curve in the upper half-plane (where $C(\eta)/\eta>0$) represents the solution for $\eta>0$ (since $C(\eta)$ must be non-negative), while its mirror image in the lower half-plane (where $C(\eta)/\eta<0$) represents it for $\eta<0$.}

\red{The fixed points of the phase plane always include two star nodes $O=(0,0)$ and $P=(1,0)$ on the horizontal axis, whose position is independent of the value of the exponent $\beta$. In addition, there is always a saddle point $S=(\beta,0)$ that lies between $O$ and $P$ for $0<\beta<1$ and to the right of $P$ for $\beta>1$, denoting a `front' where the solution is locally non-smooth. These three points $O$, $P$, and $S$ have horizontal and vertical eigendirections. The behavior of trajectories at the outer edges of the phase plane (i.e., as $U/\eta\to\pm\infty$ or as $C/\eta\to\pm\infty$) is given by three additional fixed points (labeled $M$, $N$ and $R$) that the ODE \eqref{eq:ode_g} displays when it is recast in terms of the reciprocals $\eta/U$ and $\eta/C$, as detailed in appendix \ref{sec:apdx_phase_plane}. All six fixed points are listed and classified in table \ref{tab:fixed_pts}.} 
\begin{table}
\begin{center}
\def~{\hphantom{0}}
\begin{tabular}{c c c c c}
     Fixed point & \multirow[c]{2}{2.5em}{\centering Type} & \multirow[c]{2}{4em}{\centering{Meaning}} & \multirow[c]{2}{6em}{\centering{$U(\eta)\sim$}} & \multirow[c]{2}{6em}{\centering{$C(\eta)\sim$}} \\[4pt]
     $\left(\dfrac{U}{\eta},\dfrac{C}{\eta}\right)$ & & & & \\\hline
     $O$ $(0,0)$ & SN & $|\eta|\to\infty$ & $ K\,\sgn{\eta}|\eta|^{\frac{\beta-1}{\beta}}$ & 0 \\[4pt]
      & & & $K^2\cfrac{(\beta\hspace{-0.8pt}-\hspace{-0.8pt}1)}{\beta}\,\sgn{\eta}|\eta|^\frac{\beta-2}{\beta}$ & $K|\eta|^\frac{\beta-1}{\beta}$ \\[4pt]
      & & & $K\,\sgn{\eta}|\eta|^{\frac{\beta-1}{\beta}}$ & $K'|\eta|^{\frac{\beta-1}{\beta}}$ \\\hline
     $P$ $(1,0)$ & SN {\scriptsize [$0\hspace{-1pt}<\hspace{-1pt}\beta\hspace{-1pt}<\hspace{-1pt}1$]} & $|\eta|\to\infty$ & $\eta+{K}\,\sgn{\eta}|\eta|^\frac{\beta}{\beta-1}$ & 0 \\[5pt]
     & UN {\scriptsize [$\beta\hspace{-1pt}>\hspace{-1pt}1$]} & $\eta=0$ & $\eta + \cfrac{K^2\beta}{(1\hspace{-0.8pt}-\hspace{-0.8pt}\beta)}\,\sgn{\eta}|\eta|^\frac{\beta+1}{\beta-1}$ & $K|\eta|^\frac{\beta}{\beta-1}$\\[8pt]
     & & & $\eta + K\,\sgn{\eta}|\eta|^{\frac{\beta}{\beta-1}}$ & $K'|\eta|^{\frac{\beta}{\beta-1}}$ \\\hline 
     $S$ $(\beta,0)$ & S & \multirow[b]{2}{4em}{\centering{Front at\\$\eta=\eta_f$}} & $\beta\eta\pm\eta\ln^\frac{1}{2}\hspace{-2pt}\left(\left|\cfrac{\eta}{\eta_f}\right|^{2\beta(1-\beta)}\right)$  & 0 \\[4pt]
     & & & $\beta\eta$ & $|\eta|\ln^\frac{1}{2}\hspace{-2pt}\left(\left|\cfrac{\eta_f}{\eta}\right|^{2\beta(1-\beta)}\right)$\\\hline
     $M$ $(\pm\infty,0)$ & S & $\eta=0$ & $K\,\sgn{\eta}$ & $0$\\\hline
     $N$ $(1\hspace{-1pt}-\hspace{-1pt}\beta,\pm\infty)$ & S & $\eta=0$ & $(1\hspace{-1pt}-\hspace{-1pt}\beta)\eta + \dfrac{\beta(1\hspace{-1pt}-\hspace{-1pt}\beta)(1\hspace{-1pt}-\hspace{-1pt}2\beta)}{3K^2}\eta^3$ & $K - \dfrac{\beta(1\hspace{-0.8pt}-\hspace{-0.8pt}\beta)}{2K}\eta^2$\\\hline
     $R$ $(\pm\infty,\pm\infty)$ & UN & $\eta=0$ & $K\,\sgn{\eta}$ & $K'$\\\hline
\end{tabular}
\caption{Fixed points of the ODE system given by the real and imaginary parts of \eqref{eq:ode_g}, for $\beta>0$ and $\beta\neq{1}$. SN denotes a stable node, UN an unstable node, and S a saddle. Points $M$, $N$ and $R$ are obtained as fixed points of ODE systems involving the reciprocals $\eta/U$ and $\eta/C$, as detailed in appendix \ref{sec:apdx_phase_plane}. The entries for $U(\eta)$ and $C(\eta)$ denote every possible asymptotic expansion about each fixed point, where $K$ and $K'$ are independent, real, nonzero constants of integration, and $\eta_f$ is the (real, nonzero) location of the front occurring at the saddle point $S$. When more than one entry for $U(\eta)$ and $C(\eta)$ is provided, the first \red{row corresponds to the trajectory along the horizontal eigendirection, the second row to the trajectory along the vertical eigendirection}, and the third (if provided) to generic curves \red{along any other direction}.}
\label{tab:fixed_pts}
\end{center}
\end{table}  
\red{Furthermore, the asymptotic form of the solution around each of these points can be found via linearization and is also provided in table \ref{tab:fixed_pts}, thereby listing all possible behaviors of $f(\eta)$ as $\eta\to0$ and as $|\eta|\to\infty$ for different values of $\beta$. The fact that two trajectories (one representing $\eta>0$ and another one $\eta<0$) must be `patched' at $\eta=0$ to generate a full solution $f(\eta)$ results in expansions around the origin that often involve terms like $\sgn{\eta}$ and $|\eta|$ (see table \ref{tab:fixed_pts}), which can only result in regular solutions for some specific values of $\beta$, as we show in section \ref{sec:stability}.}

While all possible similarity solutions with the correct parity can be placed in the plane, not all of them are necessarily relevant from a physical standpoint. The advantage of a phase plane formalism is that it provides a way to systematically classify all trajectories in terms of the fixed points that they connect, so that they can be identified as relevant or irrelevant. 
\begin{table}
\begin{center}
\def~{\hphantom{0}}
\begin{tabular}{c c l}
     \multirow[c]{1}{6em}{\centering{Trajectory}} & \multirow[c]{1}{10em}{\centering{Range of $\beta$}} & Physically relevant trajectory?\\\hline
     $M\to{O}$ & $\beta>0$ & No, $U(0^-)\neq{U(0^+)}$ and $C(\eta)=0$.\\[4pt]
     $R\to{O}$ & $\beta>0$ & No, $U(0^-)\neq{U(0^+)}$.\\[4pt]
     $N\to{P}$ & $0<\beta<1/2$ & No, invalid far field.\\[4pt]
     $M\to{P}$ & $0<\beta<1$ & No, $U(0^-)\neq{U(0^+)}$, $C(\eta)=0$, and invalid far field.\\[4pt]
     $R\to{P}$ & $0<\beta<1$ & No, $U(0^-)\neq{U(0^+)}$ and invalid far field.\\[4pt]
     $R\to{S}\to{O}$ & $0<\beta<1$, $\beta\neq{1/2}$ & No, $U(0^-)\neq{U(0^+)}$.\\[4pt]
     $R\to{S}\to{P}$ & $0<\beta<1$, $\beta\neq{1/2}$ & No, $U(0^-)\neq{U(0^+)}$ and invalid far field.\\[4pt]
     $N\to{S}\to{O}$ & $\beta=1/2$ & \textbf{Yes}, green trajectory in figure \ref{fig:phase_plane}.\\[4pt]
     $N\to{S}\to{P}$ & $\beta=1/2$ & No, invalid far field.\\[4pt]
     $N\to{O}$ & $\beta>1/2$ & \textbf{Yes}, purple trajectories in figure \ref{fig:phase_plane}.\\[4pt]
     $M\to{S}\to{O}$ & $\beta>1$ & No, $U(0^-)\neq{U(0^+)}$.\\[4pt]
     $P\to{S}\to{O}$ & $\beta>1$ & \textbf{Yes}, orange trajectories in figure \ref{fig:phase_plane}.\\[4pt]
     $P\to{O}$ & $\beta>1$ & \multirow[l]{3}{22em}{Depends on its direction around point $P$:\\$\bullet$ Horizontal: No, $C(\eta)=0$.\\$\bullet$ Otherwise: \textbf{Yes}, yellow trajectories in figure \ref{fig:phase_plane}.}\\[4pt]
     & & \\[8pt]
\end{tabular}
\caption{List of all possible phase-plane trajectories of the reduced similarity ODE \eqref{eq:ode_g}, for different values of the exponent $\beta$. The last column indicates whether a given trajectory represents a physically relevant solution. If the solution is classified as not relevant, the reasons are provided, based on three criteria: (i) having a nonzero concentration $C(\eta)\neq{0}$ representative of Marangoni flow, (ii) compatibility with the far-field condition \eqref{eq:far_field}, and (iii) continuity at $\eta=0$. If the trajectory is physically relevant, its color in figure \ref{fig:phase_plane} is indicated.}
\label{tab:trajectories}
\end{center}
\end{table}
We list all possible trajectories in table \ref{tab:trajectories}, where the rightmost column indicates whether the trajectory is classified as physically relevant, based on three criteria:
\begin{enumerate}
    \item Solutions must be representative of Marangoni flow. Some trajectories \red{in figure \ref{fig:phase_plane}} like $M\to{O}$, $M\to{P}$ or $P\to{O}$ are fully contained along the horizontal axis, but that implies a zero concentration $C(\eta)=0$ for all values of $\eta$, indicating that they do not represent Marangoni flow. In fact, the trajectory $P\to{O}$ along the horizontal axis for $\beta=3/2$ (figure \ref{fig:phase_plane_d}) corresponds to the real similarity solution of the inviscid Burgers equation \eqref{eq:Hopf_pde} described by \cite{Eggers2008-lk}, which appears prior to the formation of a shock and is relevant to describe other problems like gas dynamics or wave breaking.
    \item \red{Solutions must have a far-field behavior compatible with \eqref{eq:far_field}, as explained in section \ref{sec:problem_formulation_self-sim}. Accordingly, all trajectories listed in table \ref{tab:fixed_pts} with a} far field incompatible with \eqref{eq:far_field}, such as those ending at point $P$ for $0<\beta<1$, are labeled as irrelevant.
    \item Solutions must be continuous at the origin. For instance, solutions starting at points $M$ or $R$ have an odd but discontinuous velocity $U(\eta)$ at $\eta=0$, with $U(0^+)=K$ and $U(0^-)=-K$ for some real, nonzero constant $K$, as detailed in table \ref{tab:fixed_pts}. 
\end{enumerate}

Based on this classification outlined in table \ref{tab:trajectories}, we identify \emph{four} families of solutions that qualify as physically relevant. We can interpret the qualitative behavior of each of these trajectories depending on their position within the phase plane, following the discussion given in appendix \ref{sec:apdx_interpret_plane}. Trajectory $N\to{S}\to{O}$, which only exists for $\beta=1/2$, is highlighted in green in figure \ref{fig:phase_plane} and, according to appendix \ref{sec:apdx_interpret_plane}, corresponds to a `spreading' similarity solution with a forward-time definition of the similarity variable, as in equation \eqref{eq:eta_fwd}. We discuss this spreading solution, as well as its counterpart with finite diffusion, in section \ref{sec:spreading_sols}. The $P\to{O}$ (yellow in figure \ref{fig:phase_plane}) and $P\to{S}\to{O}$ (orange in figure \ref{fig:phase_plane}) trajectories are `filling' similarity solutions with a backward-time scaling as in equation \eqref{eq:eta_bwd}, which suggests they are valid immediately prior to a singularity. \red{They only hold} locally as evidenced by their far-field behavior, since these solutions only exist for $\beta>1$ and the condition \eqref{eq:far_field}, $f(\eta)\sim \red{k_{\pm\infty}}|\eta|^\frac{\beta-1}{\beta}$, then implies that they grow unbounded as $|\eta|\to\infty$. The Hilbert transform is undefined for unbounded functions so, as observed by \cite{Thess1997-qi}, similarity solutions for $\beta>1$ must be only valid locally. The last of these four families of solutions is $N\to{O}$, which exists for $\beta>1/2$, is displayed in purple in figure \ref{fig:phase_plane} and has forward-time scaling as in equation \eqref{eq:eta_fwd}. These solutions can be either spreading, for $1/2<\beta<1$, or filling, for $\beta>1$. 

Three of the identified families of \red{curves}, namely $P\to{O}$ (yellow), $P\to{S}\to{O}$ (orange), and $N\to{O}$ (purple), appear to exist for multiple values of $\beta$, which is typical of self-similar\red{ity} of the second kind \citep{Barenblatt1996-hd}. In the case of $P\to{O}$, there are even several possible solutions within the same value of $\beta$. \red{These second-kind solutions appear around finite-time singularities, and we} show in section \ref{sec:stability} how considerations about their \emph{stability} rule out many of the trajectories within these three families, leading to the only solutions that are truly obtainable in practice.

\subsection{Stability analysis} \label{sec:stability}
In order to analyze the stability of similarity solutions, we use the \emph{dynamical system} formulation \citep[see][]{Giga1985-ht,Giga1987-xg,Eggers2008-lk,Eggers2015-pz} of the inviscid Burgers equation \eqref{eq:Hopf_pde}. Instead of seeking to reduce the two physical variables $x$ and $t$ into a single similarity variable $\eta$, we use the more general change of variables
\begin{subequations} \label{eq:dyn_sys}
\begin{align}
    \psi(x,t) &= A|t-t_*|^{\beta-1}F(\eta,\tau), \label{eq:dyn_sys_F}\\
    \eta &= \sgn{t-t_*}\dfrac{x-x_*}{A|t-t_*|^\beta}, \label{eq:dyn_sys_eta}\\
    \tau &= -\ln{|t-t_*|}, \label{eq:dyn_sys_tau}
\end{align}
\end{subequations}
which, instead of reducing the original PDE to an ODE, leads to a partial differential equation for $F(\eta,\tau)$:
\begin{align} \label{eq:dyn_sys_pde}
    \pd{F}{\tau}{} = \left[(F-\beta\eta)\pd{F}{\eta}{} - (1-\beta)F\right].
\end{align}
The key of the transformation given by \eqref{eq:dyn_sys} is that the steady state of \eqref{eq:dyn_sys_pde} reduces to the similarity equation \eqref{eq:ode_f_no_diff}. Indeed, the definition \eqref{eq:dyn_sys_tau} of $\tau$ indicates that approaching the singularity time $t\to{t}_*$ corresponds to $\tau\to\infty$, leading to a steady state $\partial_\tau{F}\to{0}$ at which solutions to \eqref{eq:dyn_sys_pde} satisfy the similarity ODE \eqref{eq:ode_f_no_diff}.

The stability of each \red{of the} similarity solution\red{s} found in section \ref{sec:advect_case_phase_plane} can now be determined via linear stability analysis around $f(\eta)$. We pose a perturbation
\begin{equation} \label{eq:modes}
    F(\eta,\tau) = f(\eta) + \epp\sum_{n=0}^{\infty}b_ne^{\nu_n\tau}\phi_n(\eta) + O(\epp^2),
\end{equation}
where $\epp\ll{1}$, $\nu_n$ are the growth rates of each mode $\phi_n$, and $b_n$ are the mode amplitudes. Introducing \eqref{eq:modes} into the PDE \eqref{eq:dyn_sys_pde} we obtain at order $O(\epp)$ an eigenvalue problem
\begin{equation} \label{eq:eigen_problem}
    \LL\left[\phi_n\right] \defeq \left[\dfrac{\beta(1-\beta)}{(f-\beta\eta)}\eta+(f-\beta\eta)\td{}{\eta}{}\right]\phi_n = \nu_n\phi_n.
\end{equation}
The stability of any given similarity solution $f(\eta)$ can then be determined solving the eigenvalue problem \eqref{eq:eigen_problem} for the linear operator $\LL$, which itself depends on the similarity solution $f(\eta)$. \red{Moreover, the spectrum of $\LL$ is typically discrete given some additional conditions for the eigenfunctions $\phi_n$ \citep{Eggers2008-lk,Eggers2015-pz}.} Namely, each $\phi_n(\eta)$ must be regular at $\eta=0$, and satisfy its own quasi-stationary far-field condition
\begin{equation} \label{eq:far_field_phi}
    \phi_n(\eta) \sim \red{k_{\pm\infty}} |\eta|^\frac{\beta-1-\nu_n}{\beta} \quad \text{as } |\eta|\to\infty
\end{equation}
for some complex constant\red{s $k_{\pm\infty}$. Only those similarity solutions leading to a spectrum of $\LL$ with negative eigenvalues $\nu_n<0$ result in perturbations decaying in time and are therefore stable. However, as shown by \cite{Eggers2008-lk, Eggers2015-pz}, there is also a specific subset of non-negative eigenvalues that does not lead to instability and is instead an artifact of the continuous symmetries of the problem. Specifically, the invariance of the governing equation \eqref{eq:Hopf_pde} to shifts in space $x\to{x}+\lambda$, shifts in time $t\to{t}+\lambda$ and scalings $\psi\to\lambda\psi$, $t\to\lambda{t}$, $x\to\lambda^2x$, always leads to the three eigenvalues $\nu=\beta$, $\nu=1$, and $\nu=0$, respectively \citep[see section 3.2 in][]{Eggers2015-pz}. Accordingly, we look for solutions from the phase plane that lead to a spectrum of $\LL$ with at most these three non-negative eigenvalues, with all others being negative.}

The three families of trajectories in figure \ref{fig:phase_plane} considered here ($P\to{O}$, $P\to{S}\to{O}$ and $N\to{O}$) all have the same far-field behavior since they end at the same point. From table \ref{tab:fixed_pts}, we know that $f(\eta)\sim\red{k_{\pm\infty}}|\eta|^\frac{\beta-1}{\beta}$ as $|\eta|\to\infty$, which can be introduced in the eigenvalue problem \eqref{eq:eigen_problem} to yield the far-field behavior of the eigenfunctions:
\begin{equation} 
    \td{\phi_{n}}{\eta}{} = \left[\dfrac{\nu_{n}(f-\beta\eta)-\beta(1-\beta)\eta}{(f-\beta\eta)^2}\right]\phi_{n} \sim\left[\dfrac{\beta-1-\nu_n}{\beta\eta}\right]\phi_n \quad\text{as }|\eta|\to\infty,
\end{equation}
which leads to $\phi_n\sim\red{k_{\pm\infty}}|\eta|^\frac{\beta-1-\nu_n}{\beta}$ in the far-field, in agreement with condition \eqref{eq:far_field_phi}. \red{This means that, for these three families of trajectories, the required far-field behavior of eigenfunctions $\phi_n(\eta)$ is satisfied automatically, and the eigenvalues are then solely determined by the regularity of $\phi_n(\eta)$ at $\eta=0$, as shown in in the next three subsections.}

\subsubsection{Trajectories $P\to{O}$ (yellow in figure \ref{fig:phase_plane})} \label{sec:stability_dimple}
Table \ref{tab:fixed_pts} indicates that all $P\to{O}$ solutions that do not depart $P$ along the vertical eigendirection have expansions $U(\eta)\sim{\eta}+K\,\sgn{\eta}|\eta|^\omega$ and $C(\eta)\sim{K'}|\eta|^\omega$, where we define the exponent $\omega\defeq{\beta}/(\beta-1)$. The only possibility for $U(\eta)$ to be regular at the origin is for $\omega$ to be an odd integer (so that $\sgn{\eta}|\eta|^{\omega}=\eta^\omega$), whereas the only possibility for $C(\eta)$ to be regular is for $\omega$ to be an even integer (so that $|\eta|^{\omega}=\eta^\omega$). Since these requirements cannot be fulfilled simultaneously, these generic trajectories $P\to{O}$ can never be regular at $\eta=0$. However, the specific $P\to{O}$ trajectory that departs $P$ along the vertical eigendirection has a different expansion, given by table \ref{tab:fixed_pts} as $\red{f(\eta)\sim\eta+\ii{K}|\eta|^\omega+[K^2\beta/(\beta-1)]\,\sgn{\eta}|\eta|^{2\omega-1}}$. In that case, the expression can be regular if $\omega$ is an \emph{even} integer, turning into a true polynomial expansion. In other words, from the continuum of possible real values $\beta>1$, only a discrete set $\beta_m$ given by
\begin{equation} \label{eq:rel_beta_dimple}
    \dfrac{\beta_m}{\beta_m-1} = 2m+2, \quad\text{with }m=0,1,2,...
\end{equation}
leads to regular solutions at the origin, and only for the trajectory leaving $P$ along the vertical direction. The possible similarity exponents are then
\begin{equation} \label{eq:eigenvals_PO}
    \beta_m = \dfrac{2m+2}{2m+1} = 2, \,\dfrac{4}{3}, \,\dfrac{6}{5}, \,\dfrac{8}{7},...
\end{equation}
\red{Each value $\beta_m$ in this discrete set results in a solution $f_m(\eta)$ with an expansion $\red{f_m(\eta)\sim\eta+\ii{K}\eta^{2m+2}+K^2(2m+2)\eta^{4m+3}}$ around the origin. Moreover, each of these solutions leads to a discrete set of eigenfunctions $\phi_{mn}$ and eigenvalues $\nu_{mn}$ indexed by an integer $n$. Inserting the expansion for $f_m(\eta)$ in the eigenvalue problem \eqref{eq:eigen_problem}, we obtain}
\begin{equation} \label{eq:local_phi_dimple}
    \td{\phi_{mn}}{\eta}{} \sim \dfrac{(2m+2)-\nu_{mn}(2m+1)}{\eta}\,\phi_{mn} \quad\text{as }\eta\to{0},
\end{equation}
which implies that eigenfunctions are of the form
\begin{equation}
    \phi_{mn}(\eta) \sim {k}\,\eta^{\left[(2m+2)-\nu_{mn}(2m+1)\right]} \quad\text{as }\eta\to{0},
\end{equation}
and therefore for $\phi_{mn}$ to be smooth at the origin we require an integer exponent
\begin{equation}
    (2m+2)-\nu_{mn}(2m+1) = n, \quad\text{with }n=0,1,2,...
\end{equation}
This leads to the discrete set of eigenvalues
\begin{equation}
    \nu_{mn}=\dfrac{2m-n+2}{2m+1}, \quad\text{with }m=0,1,2,...\text{  and }n=0,1,2,...
\end{equation}
As detailed in \cite{Eggers2008-lk}, a spectrum of eigenvalues with this \red{`ladder structure' is quite general in the self-similar description of singularities}. The smallest exponent $\omega=\beta/(\beta-1)$, which is given by $m=0$, defines a `ground state' solution
\begin{equation}
    m=0 \implies \beta = \beta_0 = 2, \quad \nu_{0n}=2,\,1,\,0,\,-1,\,-2,...
\end{equation}
\red{The three non-negative eigenvalues $\nu_{00}=\beta=2$, $\nu_{01}=1$ and $\nu_{02}=0$ are an artifact of the problem symmetries and, as explained by \cite{Eggers2015-pz}, do not result in instability since their associated modes in \eqref{eq:modes} can be canceled by a shift in the constants $t_*$, $x_*$, and $A$ that enter the similarity variables \eqref{eq:dyn_sys}. All other eigenvalues are negative, so we conclude that this ground state solution $f_0(\eta)$ is stable.}

Higher values of $m$ define `excited states', such as the first two:
\begin{subequations}
\begin{align}
    m=1 \implies \beta = \beta_1 = \dfrac{4}{3}, \quad \nu_{1n}&=\dfrac{4}{3},\,1,\,\dfrac{2}{3},\,\dfrac{1}{3},\,0,\,-\dfrac{1}{3},\,-\dfrac{2}{3},... \\[5pt]
    m=2 \implies \beta = \beta_2 = \dfrac{6}{5}, \quad \nu_{2n}&=\dfrac{6}{5},\,1,\,\dfrac{4}{5},\,\dfrac{3}{5},\,\dfrac{2}{5},\,\dfrac{1}{5},\,0,\,-\dfrac{1}{5},\,-\dfrac{2}{5},... 
\end{align}
\end{subequations}
These excited states include the three eigenvalues $\nu=\beta$, $\nu=1$, $\nu=0$ that do not correspond to instability, but they also have an increasing number of other positive eigenvalues that make them unstable. Since unstable similarity solutions cannot occur in reality, these excited states are unphysical.

In summary, the only trajectory $P\to{O}$ leading to a physical solution is the one leaving $P$ along the vertical for $\beta=2$, which corresponds to the yellow trajectory highlighted with a wider streak in figure \ref{fig:phase_plane_e}. We show in section \ref{sec:filling_sols_dimple} that the similarity solution $f(\eta)$ can in this case be obtained in closed form, and that it appears when a locally depleted distribution of surfactant tends to become uniform under the action of Marangoni flow. Such a distribution, which we call a `dimple' \citep[following][]{Bickel2022-sa}, must have zero concentration $\Gamma_0(x_0)=0$ with a quadratic minimum $\Gamma_0\sim{K(x-x_0)^2}$ for some $x=x_0$. Self similarity appears prior to the time $t_*$ at which the dimple `closes', and thus we call this the `dimple closure' solution.

\subsubsection{Trajectories $P\to{S}\to{O}$ (orange in figure \ref{fig:phase_plane})} \label{sec:stability_hole}
We can proceed analogously to determine the only exponent $\beta$ that leads to stability for the $P\to{S}\to{O}$ solution. From table \ref{tab:fixed_pts}, the expansion around $P$ for trajectories departing along the horizontal eigendirection is $f(\eta)\sim\eta+K\sgn{\eta}|\eta|^\omega$, with $\omega\defeq\beta/(\beta-1)$. This solution can only be regular at the origin if the exponent $\omega$ is \emph{odd}, excluding $\omega=1$ since it cannot be achieved for any finite $\beta$. We then have the discrete sequence
\begin{equation}
    \dfrac{\beta_m}{\beta_m-1} = 2m+3, \quad\text{with }m=0,1,2,...\,\,,
\end{equation}
which results in exponents given by
\begin{equation}
    \beta_m = \dfrac{2m+3}{2m+2} = \dfrac{3}{2},\,\dfrac{5}{4},\,\dfrac{7}{6},\,\dfrac{9}{8},...
\end{equation}
This set of values leads to local expansions $f(\eta) \sim \eta + K\eta^{2m+3}$ around $\eta=0$. Introducing this expansion for $f(\eta)$ in the eigenvalue problem \eqref{eq:eigen_problem}, we obtain
\begin{equation} \label{eq:local_phi_hole}
    \td{\phi_{mn}}{\eta}{} \sim \dfrac{(2m+3)-\nu_{mn}(2m+2)}{\eta}\,\phi_{mn} \quad\text{as }\eta\to{0},
\end{equation}
which leads to eigenfunctions $\phi_{mn}\sim{k}\,\eta^{\left[(2m+3)-\nu_{mn}(2m+2)\right]}$ locally around $\eta=0$. The functions $\phi_{mn}$ are then smooth only if
\begin{equation}
    (2m+3) - \nu_{mn}(2m+2) = n, \quad\text{with }n=0,1,2,...
\end{equation}
This means that the discrete sequence of eigenvalues for the $P\to{S}\to{O}$ trajectory is
\begin{equation}\label{eq:spectrum_hole}
    \nu_{mn} = \dfrac{2m-n+3}{2m+2}, \quad\text{with }m=0,1,2,...\text{  and }n=0,1,2,...
\end{equation}
This spectrum of eigenvalues is identical to that of the \emph{real} inviscid Burgers equation, for which \cite{Eggers2008-lk} show there exists a similarity solution of the second kind with $\beta=3/2$ immediately prior to the formation of a shock. This should not come as a surprise, since that (real) similarity solution is simply the $P\to{O}$ trajectory along the \emph{horizontal axis} in figure \ref{fig:phase_plane_e}. This $P\to{O}$ solution and the $P\to{S}\to{O}$ solution both depart $P$ along the horizontal eigendirection, so they have the same leading-order structure around $\eta=0$ and thus the same spectrum. Similar to the previous case of section \ref{sec:stability_dimple}, the eigenvalues \eqref{eq:spectrum_hole} lead to states of the form
\begin{subequations}
\begin{align}
        m&=0 \implies \beta = \beta_0 = \dfrac{3}{2}, \quad \nu_{0n}=\dfrac{3}{2},\,1,\,\dfrac{1}{2},\,0,\,-\dfrac{1}{2},\,-1,...\\[5pt]
        m&=1 \implies \beta = \beta_1 = \dfrac{5}{4}, \quad \nu_{1n}=\dfrac{5}{4},\,1,\,\dfrac{3}{4},\,\dfrac{1}{2},\,\dfrac{1}{4},\,0,\,-\dfrac{1}{4},\,-\dfrac{1}{2},...\\[5pt]
        m&=2 \implies \beta = \beta_2 = \dfrac{7}{6}, \quad \nu_{2n}=\dfrac{7}{6},\,1,\,\dfrac{5}{6},\,\dfrac{2}{3},\,\dfrac{1}{2},\,\dfrac{1}{3},\,\dfrac{1}{6},\,0,\,-\dfrac{1}{6},\,-\dfrac{1}{3},...
    \end{align}
\end{subequations}
\red{All states contain the eigenvalues $\nu=\beta$, $\nu=1$ and $\nu=0$ that do not lead to instabilities, but also have other positive eigenvalues that seemingly imply that no stable similarity solutions exist. However, in the particular case of the ground-state solution $m=0$, the additional positive eigenvalue $\nu=1/2$ does not necessarily lead to instability either, as shown by \cite{Eggers2008-lk} in the case of the real solution. This can be shown particularizing the expansion \eqref{eq:modes} at the position $x=x_*$ (or $\eta=0$) of the singularity,}
\begin{equation}
    \red{\left.\pd{F_0}{\eta}{2}\right|_{\eta=0} = f_0''(0) + \, \epp\left[{b_{02}}\,e^{\tau/2}\phi_{02}''(0) + \sum_{\substack{n=0 \\ n\neq 2}}^{\infty}b_{0n}\,e^{\nu_{0n}\tau}\phi_{0n}''(0)\right] + O(\epp^2).}
    \label{eq:modes_particularized}
\end{equation}
\red{Since we have deduced that for the ground state $m=0$ we have $f_0\sim\eta+K\eta^3$ and $\phi_{0n}(0)\sim{k}\,\eta^n$ around the origin, then we have that $f_0''(0)=0$ and $\phi_{0n}''(0)=0$ for all modes with $n\neq{2}$, canceling out the first and third terms on the right-hand side of \eqref{eq:modes_particularized}. Also, we have that $\phi_{02}''(0)\neq{0}$ in the second term above. However, in the particular case where the second derivative of the solution is zero at the position of the singularity, i.e., $\partial_{xx}\psi(x_*,t)=0$, we also have $\partial_{\eta\eta}F(0,\tau)=0$ and therefore the amplitude $b_{02}$ must be zero. In that case, $b_{02}=0$ cancels the $n=2$ mode altogether and the positive eigenvalue $\nu=1/2$ is irrelevant in the stability of the solution.} As a consequence, the ground-state $m=0$ is stable for the particular set of initial conditions that lead to $\partial_{xx}\psi=0$ at the position of the singularity. In the case of the real similarity solution studied by \cite{Eggers2008-lk}, one can show that this is always satisfied \red{in a suitable frame of reference} for solutions that develop a shock (see appendix \ref{sec:apdx_closure_time}). However, the condition is not necessarily satisfied in the complex case. In fact, we show in section \ref{sec:filling_sols} that only if the initial distribution of surfactant is locally depleted with $\Gamma_0(x_0)=0$ for some $x=x_0$, and is \emph{also} `flatter' than quadratic (i.e., if $\Gamma_0''(x_0)=0$), then the condition $\partial_{xx}\psi(x_*,t)=0$ is fulfilled and the self-similar solution before the singularity is then $P\to{S}\to{O}$ with $\beta=3/2$. We call these flatter surfactant profiles `holes', in order to distinguish them from dimples with a sharper quadratic minimum. 

In conclusion, $\beta=3/2$ is the only exponent leading to stability for $P\to{S}\to{O}$, corresponding to the trajectory highlighted in figure \ref{fig:phase_plane_d}. Only filling solutions that have $\partial_{xx}\psi=0$ at the singularity, which we call `holes', can lead to this similarity solution, which we correspondingly call the `hole closure' solution. We show in section \ref{sec:filling_sols_hole} that $f(\eta)$ can in this case also be obtained in closed form.

\subsubsection{Trajectories $N\to{O}$ (purple in figure \ref{fig:phase_plane})} \label{sec:stability_leveling}
The similarity solution $N\to{O}$ has a functional form $\red{f(\eta)\sim\ii{K}+(1-\beta)\eta}$ around $\eta=0$ (see table \ref{tab:fixed_pts}), appearing to always be smooth at the origin $\eta=0$ independently of $\beta$. Furthermore, inserting it into the eigenvalue problem \eqref{eq:eigen_problem}, we obtain
\begin{equation}
    \td{\phi_n}{\eta}{}\sim\ii\nu_n\phi_n \quad\text{as }\eta\to{0},
\end{equation}
which always leads to smooth eigenfunctions $\phi_n\sim{k}e^{\ii\nu_n\eta}$ around $\eta=0$. The absence of an evident sequence of discrete solutions or eigenfunctions based on regularity suggests that the behavior of the $N\to{O}$ solution could be governed by a more complicated continuum of possible similarity solutions \citep[as in][]{Eggers2000-ep}. 

Analyzing every possible exponent for the $N\to{O}$ solution is beyond the scope of this paper, but we will consider the two specific cases of $\beta=3/2$ and $\beta=2$. \red{Since, according to appendix \ref{sec:apdx_interpret_plane},} $N\to{O}$ can be interpreted as a forward-time solution happening subsequent to a singularity, it must appear immediately after either the `dimple closure' solution or the `hole closure' solution, since both occur immediately before the singularity. Furthermore, since the far-field scaling of solutions is intimately linked to the similarity exponents through the condition \eqref{eq:far_field}, the exponent of the pre-singularity solution fixes the exponent of the post-singularity $N\to{O}$ solution\red{,} otherwise the far-field behavior of the solution would change instantly \red{at $t=t_*$}. We can then ensure that $\beta=3/2$ and $\beta=2$ are possible exponents for the $N\to{O}$ solution, as we confirm in section \ref{sec:filling_sols}. Following the naming convention used by \cite{Zheng2018-bu} for capillary films, we use the term `leveling' for these two solutions in which the concentration levels towards $\Gamma\to{1}$, as opposed to the pre-singularity `closure' solutions where the concentration remains $\Gamma=0$ at the point of the singularity. Specifically, we call the $N\to{O}$ solution with $\beta=2$ the `dimple leveling' solution \red{(highlighted in figure \ref{fig:phase_plane_e})}, since it follows the `dimple closure' solution, whereas the $N\to{O}$ solution with $\beta=3/2$ is labeled as the `hole leveling' solution \red{(highlighted in figure \ref{fig:phase_plane_d})} since it comes after the `hole closure' solution. These solutions are obtained in closed form in sections \ref{sec:filling_sols_dimple} and \ref{sec:filling_sols_hole}, respectively.

\section{Spreading solutions} \label{sec:spreading_sols}
In the case of a spreading pulse, depicted in figure \ref{fig:problem_setup_spreading}, the total mass of (insoluble) surfactant is, in general, conserved and imposed by the initial profile $\Gamma_0(x)$. This can be realized by direct integration of equation \eqref{eq:surfactant_pde} in $x$, as we show in appendix \ref{sec:apdx_invariants}. We define the total (dimensionless) mass $M_0$ as
\begin{equation} \label{eq:def_M0}
    M_0 \defeq \int_{-\infty}^{\infty}\Gamma(x\red{, t})\dd{x} = \int_{-\infty}^{\infty}\Gamma_0(x)\dd{x},
\end{equation}
which, upon substitution of the similarity ansatzes \eqref{eq:ansatz_eta} and \eqref{eq:Gamma_f}, and using \red{$A=B$} and a forward-time description $t>t_*$ as discussed in section \ref{sec:problem_formulation_self-sim}, leads to
\begin{equation}
    M_0 = A^2(t-t_*)^{\alpha+\beta}\int_{-\infty}^{\infty}C(\eta)\dd{\eta}.
    \label{eq:integral_constraint_exponents}
\end{equation}
The above relation is only compatible with self-similar behavior if the additional requirement $\alpha+\beta=0$ is met, in which case the solution $f(\eta)$ must also satisfy
\begin{equation}
    \int_{-\infty}^{\infty}C(\eta)\dd{\eta} = \dfrac{M_0}{A^2}.
    \label{eq:integral_constraint_pulse}
\end{equation}

\subsection{Advection-dominated limit} \label{sec:spreading_sols_no_diff}
For the limit of zero diffusion $Pe_s^{-1}=0$, the requirement $\alpha=\beta-1$ from the similarity ODE \eqref{eq:ode_f_exponents}, combined with $\alpha=-\beta$ from the integral constraint \eqref{eq:integral_constraint_exponents}, leads to $\alpha=-1/2$ and $\beta=1/2$. Since the exponents can be fixed a priori only from dimensional analysis, this solution displays self-similarity of the first kind \citep[][]{Barenblatt1996-hd}. This solution corresponds to the $N\to{S}\to{O}$ trajectory in figure \ref{fig:phase_plane_b}, as shown in section \ref{sec:advect_case}.

For these particular values of the exponents $\alpha$ and $\beta$, an explicit solution can be obtained from the implicit relationship \eqref{eq:implicit_sol_f}. For $\beta=1/2$, equation \eqref{eq:implicit_sol_f} yields
\begin{equation}
    f^2 - \eta{f} + k_\pm = 0. 
    \label{eq:polyn_f_pulses}
\end{equation}
\red{The case of a pulse solution requires $f(0)=\ii\,{C}(0)$ to be imaginary, which applied to \eqref{eq:polyn_f_pulses} leads to real integration constants $k_\pm$. Added to the required symmetry condition  $k_-=\overline{k}_+$ \eqref{eq:k_condition}, this means that we can consider only a single real constant $k=k_+=k_-$}. Furthermore, and since solutions to the similarity ODE \eqref{eq:ode_f_no_diff} are defined only up to a rescaling of $f$ and $\eta$, we choose $k=1$ to fix $\red{f(0)=\ii}$ and, by extension, $C(0)=1$. The quadratic equation can be solved, leading to
\begin{equation}
    f(\eta)=\dfrac{1}{2}\left(\eta\pm\sqrt{\eta^2-4}\right),
\end{equation}
\red{where the root symbol is always taken to indicate the principal root. The choice of sign must be made at each value of $\eta$ to ensure that the concentration $\im{f}$ remains positive and that the far-field complies with condition \eqref{eq:far_field}. These conditions lead} to a particularly compact final form of the solution
\begin{equation} \label{eq:f_pulses}
    \red{f(\eta) = \dfrac{1}{2}\left(\eta+\sqrt{2-\eta}\sqrt{-2-\eta}\right).}
\end{equation}
The complex form \eqref{eq:f_pulses} can be split into its real and imaginary parts to yield the similarity solutions $U(\eta)$ and $C(\eta)$ separately, which can be defined piecewise as
\begin{subequations} \label{eq:f_pulses_ReIm}
    \begin{align}
        C(\eta)&=
        \begin{cases}
            \cfrac{1}{2}\sqrt{4-\eta^2}\hphantom{\dfrac{1}{2}\left[\eta+\right]} & \text{if }|\eta|\leq{2},\\[8pt]
            0 & \text{if }|\eta|\geq{2}.
        \end{cases}\\[8pt]
        U(\eta)&=
        \begin{cases}
            \dfrac{\eta}{2} & \text{if }|\eta|\leq{2},\\[8pt]
            \dfrac{1}{2}\,\sgn{\eta}\left[|\eta| - \sqrt{\eta^2-4}\right] & \text{if } |\eta|\geq{2}.
        \end{cases}
    \end{align}
\end{subequations}
Note that, \red{since $\int_{-\infty}^\infty{C}(\eta)\dd\eta=\pi$, in order to} satisfy the integral constraint \eqref{eq:integral_constraint_pulse} the constant $A$ in the similarity ansatz \eqref{eq:ansatz} must be chosen as
\begin{equation}
    A=\sqrt{\dfrac{M_0}{\pi}}.
\end{equation}

\begin{figure}
    \centering 
    \subfloat{\includegraphics[]{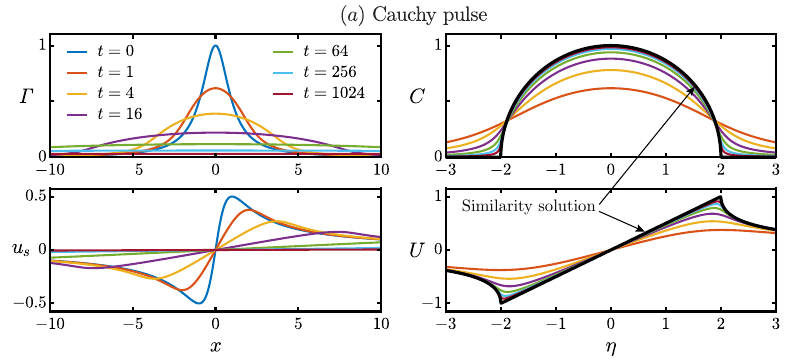}\label{fig:spreading_no_diff_cauchy}} \\[-2pt]
    \subfloat{\includegraphics[]{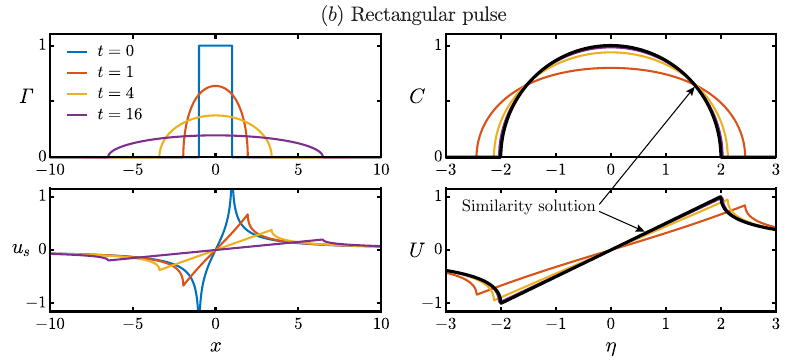}\label{fig:spreading_no_diff_rectangular}} \\[-2pt]
    \subfloat{\includegraphics[]{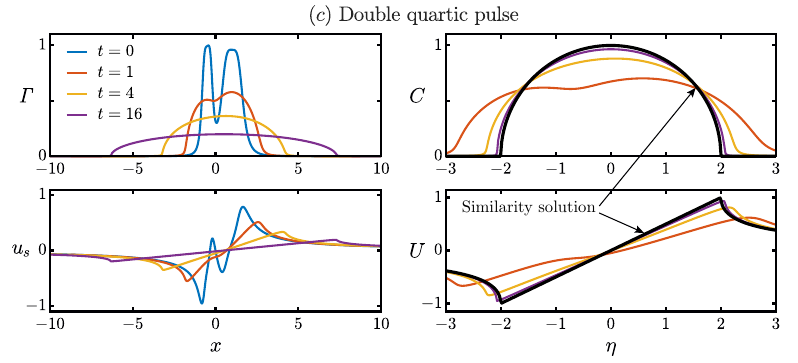}\label{fig:spreading_no_diff_double}}
    \caption{Spreading solutions with $Pe_s^{-1}=0$, for initial profiles of surfactant given by $(a)$ a Cauchy pulse, $(b)$ a rectangular pulse, and $(c)$ a double quartic pulse, with their functional forms given in appendix \ref{sec:apdx_hilbert_transforms}. For each example, left panels show the concentration $\Gamma(x,t)$ and interfacial velocity $u_s(x,t)$ obtained through the exact solution of \eqref{eq:Hopf}, while right \red{panels show the same data rescaled in similarity variables (color curves), superimposed to the similarity solution \eqref{eq:f_pulses_ReIm} (black curves) valid at long times $t\gg{1}$.} For the double quartic pulse, the reference position in \eqref{eq:ansatz} is $x_*=1/2$; in all other examples $x_*=t_*=0$.}
    \label{fig:spreading_no_diff}
\end{figure}

Figure \ref{fig:spreading_no_diff}
displays three distinct spreading pulses of surfactant (whose initial profiles $\Gamma_0(x)$ can be found in appendix \ref{sec:apdx_hilbert_transforms}) obtained solving the inviscid Burgers equation \eqref{eq:Hopf_pde} via the method of characteristics \citep[][]{Crowdy2021-vy}. At large times $t\gg{1}$, the curves are shown to collapse onto the similarity solution \eqref{eq:f_pulses}, which is equivalent to the one originally identified by \cite{Thess1996-bc} through different methods. As pointed out by \cite{Bickel2022-sa}, a Cauchy pulse (figure \ref{fig:spreading_no_diff_cauchy}), which decays as $\Gamma\sim{x}^{-2}$ in the far field, requires very large times of order $t={O}(10^3)$ to become visually indistinguishable from the similarity solution. However, figures \ref{fig:spreading_no_diff_rectangular} and \ref{fig:spreading_no_diff_double} illustrate how pulses with a faster decay or compact support require much shorter times, $t={O}(10)$ to converge to \eqref{eq:f_pulses}.

It is worth noting that the initial profile $\Gamma_0(x)$ of the double quartic pulse of figure \ref{fig:spreading_no_diff_double} is \emph{asymmetric}, and therefore the center $x_*$ of the distribution at long times is nonzero. We show in appendix \ref{sec:apdx_invariants} that the first moment of the surfactant distribution
\begin{equation} \label{eq:def_M01}
    M_1 \defeq \int_{-\infty}^{\infty}x\Gamma(x,t)\dd{x}=\int_{-\infty}^{\infty}x\Gamma_0(x)\dd{x}
\end{equation}
is a conserved invariant of the problem, provided the above integral exists. This leads to a straightforward definition of the reference position $x_*$ for pulses, namely
\begin{equation} \label{eq:x_star_pulses}
    x_* = \dfrac{\int_{-\infty}^{\infty}x\Gamma_0(x)\dd{x}}{\int_{-\infty}^{\infty}\Gamma_0(x)\dd{x}} = \dfrac{M_1}{M_0}.
\end{equation}
In the case of figure \ref{fig:spreading_no_diff_double}, we have that $x_*=1/2$ (see appendix \ref{sec:apdx_hilbert_transforms}), as can be evidenced by the shifted pulse at long times in the top left panel.

\subsection{General case with finite diffusion} \label{sec:spreading_sols_general}
For general values of $Pe_s^{-1}>0$, equation \eqref{eq:ode_f_exponents} \red{requires both exponents to be fixed} $\alpha=-1/2$ and $\beta=1/2$. \red{These values happen to also be compatible with the} additional requirement $\alpha+\beta=0$ from the integral constraint \red{\eqref{eq:integral_constraint_exponents}}, illustrating that this more general case \red{with diffusion} also displays self-similar solutions of the first kind. The governing ODE given by \eqref{eq:ode_f_full} can be integrated directly, leading to 
\begin{equation} \label{eq:riccati_eqn}
    \td{f}{\eta}{} = k_1 - \eta{f} + f^2,
\end{equation}
where the constant $k_1$ must be real since, \red{for pulses}, \red{$f(0)=\ii\,C(0)$} is imaginary and $\red{f'(0)=U'(0)}$ is real \red{by symmetry}. The far-field condition \eqref{eq:far_field}, which in this case with $\beta=1/2$ translates to $f\sim\red{{k}_{\infty}}\,\eta^{-1}$ as $\eta\to\infty$, can be introduced in \eqref{eq:riccati_eqn} to obtain that $k_1=\red{{k}_{\infty}}$, meaning that the constant $k_1$ in \eqref{eq:riccati_eqn} is simply the prefactor in the leading order far-field behavior of $f(\eta)$. This constant can be obtained realizing that, at the initial time $t\to{t}_*$, the solution must converge to a single Dirac distribution of surfactant with mass $M_0$ centered at $x_*$, and with an interfacial velocity that must be the Hilbert transform of that Dirac distribution. This can be stated mathematically as
\begin{equation}
    A|t-t_*|^{-1/2}f(\eta) \sim \red{\hil{M_0\delta(x-x_*)} + \ii\,M_0\delta(x-x_*)} \quad\text{as }t\to{t}_*.
\end{equation}
Using the linearity of the Hilbert transform, the fact that $\hil{\delta(x)}=(\pi{x})^{-1}$ \citep[][]{King2009-bq} and the rescaling property $\delta(Kx)=\delta(x)/K$ of the Dirac distribution, we obtain
\begin{equation}
    A|t-t_*|^{-1/2}f(\eta) \sim \dfrac{M_0}{A}|t-t_*|^{-1/2}\left[\red{\dfrac{1}{\pi\eta}+\ii\,\delta(\eta)}\right] \quad\text{as }\eta\to\infty,
\end{equation}
and, since we had chosen $A=\sqrt{2/Pe_s}$ in section \ref{sec:problem_formulation_self-sim}, this means that
\begin{equation}
    f(\eta) \sim \dfrac{M_0Pe_s}{2\pi}\,\eta^{-1} \quad\text{as }\eta\to\infty.
\end{equation}
Hence, the integration constant \red{in \eqref{eq:riccati_eqn}} must be 
\begin{equation} \label{eq:def_k_1}
    k_1=\red{{k}_{\infty}}=\dfrac{M_0Pe_s}{2\pi}.
\end{equation}

\begin{figure}
    \centering 
    \subfloat{\includegraphics[]{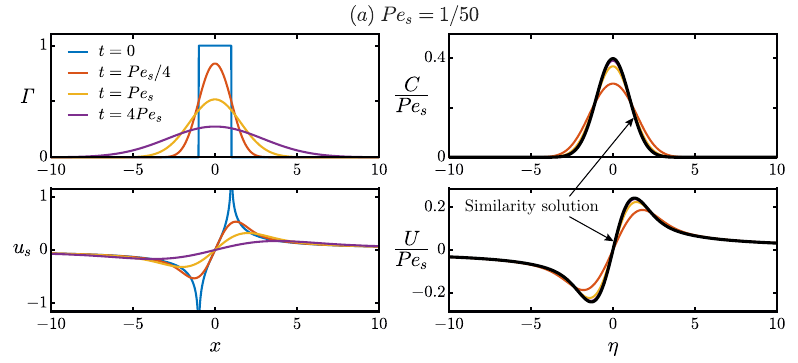}\label{fig:spreading_diff_lowPe}} \\[-2pt]
    \subfloat{\includegraphics[]{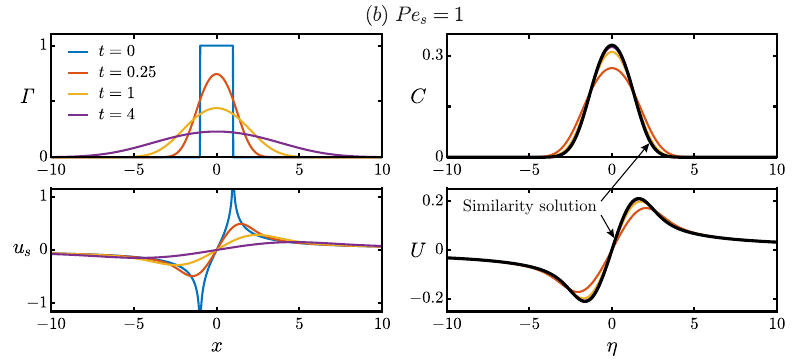}\label{fig:spreading_diff_onePe}} \\[-2pt]
    \subfloat{\includegraphics[]{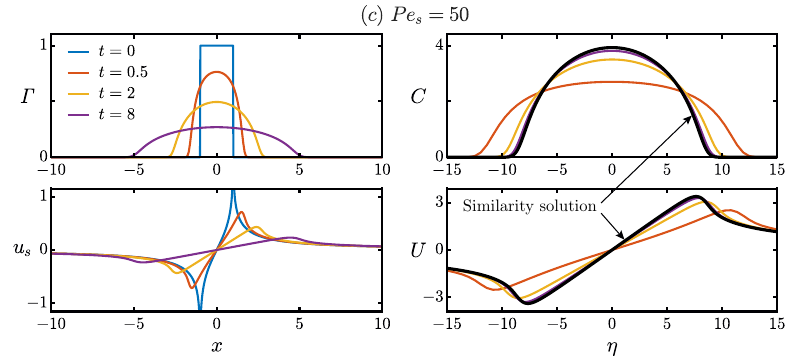}\label{fig:spreading_diff_highPe}}
    \caption{Spreading solutions for an initially rectangular pulse of surfactant, with a functional form given in appendix \ref{sec:apdx_hilbert_transforms}, for $(a)$ $Pe_s=1/50$, $(b)$ $Pe_s=1$, and $(c)$ $Pe_s=50$. For each example, left panels show the concentration $\Gamma(x,t)$ and interfacial velocity $u_s(x,t)$ obtained through the exact solution of \eqref{eq:Burgers}, while right \red{panels show the same data rescaled in similarity variables (color curves), superimposed to the similarity solution \eqref{eq:f_pulses_diff} (black curves) valid at long times $t\gg{1}$.} In all examples, $x_*=t_*=0$.}
    \label{fig:spreading_diff}
\end{figure}

The ODE \eqref{eq:riccati_eqn} can be further integrated by noting that it is a Riccati equation, which can be solved with the change of dependent variable
\begin{equation} \label{eq:riccati_change_vari}
    f = - \dfrac{1}{h}\td{h}{\eta}{} = - \td{}{\eta}{}\Ln(h),
\end{equation}
\red{with $\Ln(\,\,)$ the principal value of the complex logarithm. This} leads to a linear equation 
\begin{equation}\label{eq:riccati_ode_h}
    \td{h}{\eta}{2} + \eta\td{h}{\eta}{} + k_1{h} = 0.
\end{equation}
Note the analogy between the Cole-Hopf transformation used to  linearize Burgers equation directly \citep[][]{Crowdy2021-vy,Bickel2022-sa} and the change of variables \eqref{eq:riccati_change_vari} to linearize the similarity ODE \eqref{eq:riccati_eqn}. The solution to equation \eqref{eq:riccati_ode_h} is
\begin{equation}\label{eq:h_sol}
    h(\eta) = k_2\,\hyperM{\dfrac{k_1}{2}}{\dfrac{1}{2}}{-\dfrac{\eta^2}{2}} + k_3\,\eta\,\hyperM{\dfrac{1}{2}+\dfrac{k_1}{2}}{\dfrac{3}{2}}{-\dfrac{\eta^2}{2}},
\end{equation}
where $\hyperM{a}{b}{z}$ is Kummer's confluent hypergeometric function \citep[][]{Olver2010-qx} and $k_2$, $k_3$ are complex integration constants. Since, as evidenced by the change of variables \eqref{eq:riccati_change_vari}, the solution $f(\eta)$ is independent of any rescalings of $h(\eta)$ with a complex constant, we can set $k_3=1$ without any loss of generality. The remaining constant $k_2$ indicates the value of the concentration $C$ at the origin, since
\begin{equation}
    f(0) = -\dfrac{1}{h(0)}\left.\td{h}{\eta}{}\right|_{\eta=0} = -\dfrac{1}{k_2},
\end{equation}
which highlights that $k_2$ must be imaginary with $\red{\im{k_2}>0}$. The value of $k_2$ can be obtained imposing the integral constraint given by \eqref{eq:integral_constraint_pulse}, namely
\begin{equation}
    \int_{-\infty}^{\infty}f\dd{\eta} =-\int_{-\infty}^{\infty}\td{}{\eta}{}\Ln(h) \dd{\eta} = \ii\,\dfrac{M_0Pe_s}{2},
\end{equation}
which can be simplified as
\begin{equation}
    \lim_{\eta\to\infty}\Bigl[\Ln(h(\eta))-\Ln(h(-\eta))\Bigr] = \red{-}\ii\,\dfrac{M_0Pe_s}{2}.
\end{equation}
Since \red{$k_2$ is imaginary, $k_1$ and $k_3$ are real, and} the hypergeometric functions in \eqref{eq:h_sol} are a function only of $\eta^2$, we can conclude that \red{the parity of $h$ must be} $h(-\eta)=-\bar{h}(\eta)$. Using this identity, and splitting $\Ln(z)=\ln|z|+\ii\Arg(z)$, we get
\begin{equation}
    \red{\lim_{\eta\to\infty}\Arg(h(\eta)) = - \dfrac{M_0Pe_s}{4} \pm \dfrac{\pi}{2}}
\end{equation}
which, after considering the far-field behavior of $\hyperM{a}{b}{z}$ \citep[][]{Olver2010-qx} and the fact that $k_1 = M_0Pe_s/(2\pi)$, leads to 
\begin{equation} \label{eq:def_k_2}
    \red{k_2 = \ii \, \dfrac{\sqrt{2}}{2}\dfrac{\Gam{\dfrac{M_0Pe_s}{4\pi}}}{\Gam{\dfrac{1}{2}+\dfrac{M_0Pe_s}{4\pi}}}},
\end{equation}
where $\Gam{\,\,}$ is the gamma function and should not be confused with the surfactant concentration $\Gamma$, which is italicized throughout the paper. Undoing the change of variables \eqref{eq:riccati_change_vari}, the final form of the similarity solution is
\begin{subequations} \label{eq:f_pulses_diff}
    \begin{equation}
        \red{f(\eta) =\tfrac{ - 6\,\sGam{\frac{1}{2}+\zeta}\hyperM{\frac{1}{2}+\zeta}{\frac{3}{2}}{-\frac{\eta^2}{2}} + 6\sqrt{2}\,\ii\,\eta\,\sGam{1+\zeta}\hyperM{1+\zeta}{\frac{3}{2}}{-\frac{\eta^2}{2}} + 4\,\eta^2\,\sGam{\frac{3}{2}+\zeta}\hyperM{\frac{3}{2}+\zeta}{\frac{5}{2}}{-\frac{\eta^2}{2}} }{ 3\sqrt{2}\,\ii\,\sGam{\zeta}\hyperM{\zeta}{\frac{1}{2}}{-\frac{\eta^2}{2}} + 6\eta\,\sGam{\frac{1}{2}+\zeta}\hyperM{\frac{1}{2}+\zeta}{\frac{3}{2}}{-\frac{\eta^2}{2}} }},
    \end{equation}
where we have defined
    \begin{equation}
        \zeta \defeq \dfrac{M_0Pe_s}{4\pi}.
    \end{equation}
\end{subequations}
The solution given by \eqref{eq:f_pulses_diff} is equivalent to the fundamental solution derived by \cite{Bickel2022-sa} using the Cole-Hopf transformation with a Dirac distribution as the initial condition. Figure \ref{fig:spreading_diff}
displays an initially rectangular pulse of surfactant spreading following Burgers equation \eqref{eq:Burgers_pde}, for different values of the surface P\'eclet number. When advection is negligible and $Pe_s\ll{1}$, the solution quickly becomes Gaussian in shape, converging towards the fundamental solution of the (linear) diffusion equation. This occurs on times of order $t= O(Pe_s)$, since at small $Pe_s$ the dominant balance in equation \eqref{eq:Burgers_pde} modifies the characteristic timescale, which we had assumed to be set by advection in section \ref{sec:problem_formulation}. As $Pe_s$ increases and reaches an advection-dominated regime $Pe_s\gg{1}$, the solution changes shape, resembling the semicircular surfactant profile of the purely advective solution \eqref{eq:f_pulses} of the previous subsection. 

\section{Filling solutions} \label{sec:filling_sols}
In the case of filling solutions, sketched in figure \ref{fig:problem_setup_filling}, there is no conserved mass of surfactant since the integral of $\Gamma_0(x)$ diverges as $\Gamma_0\to{1}$ in the far field. This leads to self-similar solutions of the second kind \citep{Barenblatt1996-hd}, where the exponent $\beta$ cannot be determined from dimensional considerations, but is instead given by the stability criteria presented in section \ref{sec:stability}. Furthermore, the scaling constant $A$ is in this case dependent on the local properties of initial conditions, and can only be either computed numerically or calculated if a full solution $\psi(x,t)$ to \eqref{eq:Hopf} can be obtained explicitly. 

The four filling solutions identified in section \ref{sec:advect_case} hold only locally, either before or after a reference \red{`closure'} time $t_*$ at which the derivative of the solution is singular. This singular behavior has only been observed \citep{Thess1997-qi,Crowdy2021-vy,Bickel2022-sa} when the initial distribution of surfactant is zero $\Gamma_0(x)=0$ somewhere along the real line. \red{This fact allows to calculate, using the method of characteristics, the time $t_*$, position $x_*$, and velocity $u_*$ of the \red{singular point} \emph{a priori} from the initial surfactant profile $\Gamma_0(x)$, as we illustrate in appendix \ref{sec:apdx_closure_time}. While symmetric surfactant profiles result in a static singularity $u_*=0$ \citep[as in][]{Thess1997-qi,Crowdy2021-vy,Bickel2022-sa}, note that the (constant) velocity of the moving point at which the singularity develops can in general be nonzero for asymmetric profiles.}

\red{The method of characteristics can similarly} be used to illustrate which initial conditions \red{are} classified as `dimples', leading to $\beta=2$, and which ones to `holes', resulting in $\beta=3/2$. As \red{mentioned} in section \ref{sec:stability}, the key distinction between initial profiles $\Gamma_0(x)$ that lead to one or the other similarity solution is the second derivative of the solution at the singularity, which we calculate \red{in a frame of reference moving with the singular point}. To that end, we first define the position of the \red{singular point} as $x_s(t)\defeq{x}_*-u_*(t_*-t)$. Then, we \red{use the method of characteristics to obtain $\partial_{xx}\psi(x,t)$, which is given in equation \eqref{eq:dxx_char}, and particularize it at $x_s(t)$} to obtain
\begin{equation}
    \pd{\psi}{x}{2}(x_s(t),t) = \dfrac{\psi_0''(x_*-t_*u_*)}{\left(1+t\,\psi_0'(x_*-t_*u_*)\right)^3}.
\end{equation}
In the case of real solutions, it can be shown that $\psi_0''(x_*-t_*\,u_*)=0$ (see appendix \ref{sec:apdx_closure_time}), and therefore $\partial_{xx}\psi=0$ for all times at the (moving) point of the shock. However, complex solutions can lead to either $\psi_0''(x_*-t_*u_*)\neq{0}$, in which case we define the initial distribution $\Gamma_0(x)$ as a `dimple', or to $\psi_0''(x_*-t_*u_*)={0}$, in which case we define it as a `hole'. Each of these two cases lead to a different similarity solution and are therefore treated separately in the next two subsections.

\subsection{Dimple solutions} \label{sec:filling_sols_dimple}
\begin{figure}
    \centering 
    \subfloat{\includegraphics[]{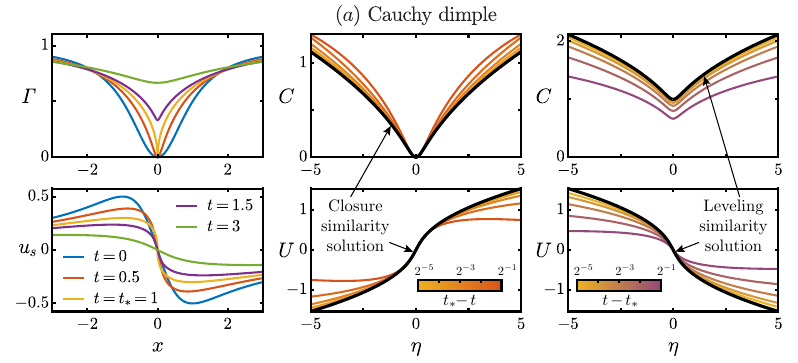}\label{fig:dimple_cauchy}} \\[-2pt]
    \subfloat{\includegraphics[]{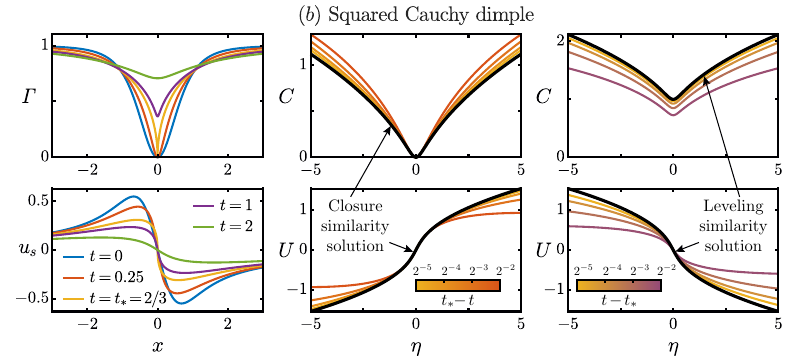}\label{fig:dimple_squared_cauchy}} \\[-2pt]
    \subfloat{\includegraphics[]{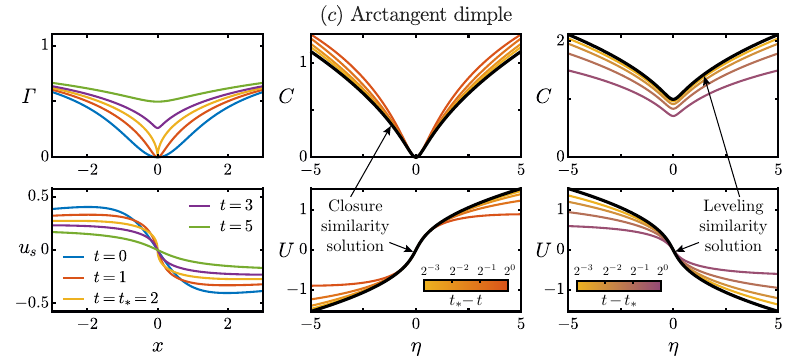}\label{fig:dimple_arctan}} \\[-2pt]
    \caption{Filling solutions with $Pe_s^{-1}=0$, for initial `dimple' distributions of surfactant given by $(a)$ a Cauchy dimple, $(b)$ a squared Cauchy dimple, and $(c)$ an arctangent dimple, with their functional forms given in appendix \ref{sec:apdx_hilbert_transforms}. For each example, the left column shows the concentration $\Gamma(x,t)$ and interfacial velocity $u_s(x,t)$ obtained through the exact solution of \eqref{eq:Hopf}. The middle and right columns show exact solutions rescaled in similarity variables (color curves), superimposed to the closure \eqref{eq:f_dimple_closure_ReIm} and leveling \eqref{eq:f_dimple_leveling_ReIm} similarity solutions (black curves) valid prior and subsequent to the singularity, respectively. In all examples, $x_*=0$.}
    \label{fig:dimple}
\end{figure}

The first dimple distribution we consider is $\Gamma_0(x)=x^2/(1+x^2)$ (which we call the `Cauchy dimple') since it has already been studied by \cite{Crowdy2021-vy} and \cite{Bickel2022-sa}. It has a quadratic minimum $\Gamma_0(x)\sim{x}^2$ around $x=0$ and, since it is a symmetric distribution, it follows from the method of characteristics (see appendix \ref{sec:apdx_closure_time}) that $u_*=x_*=0$ and $t_*=1$. The exact evolution of a Cauchy dimple is displayed in the left column of figure \ref{fig:dimple_cauchy},
with the surfactant concentration reaching a cusp-like singularity at $t_*$ and $x_*$. The left column of figures \ref{fig:dimple_squared_cauchy} and \ref{fig:dimple_arctan} displays the evolution of other profiles $\Gamma_0(x)$ with different functional forms (detailed in appendix \ref{sec:apdx_hilbert_transforms}), but always with a quadratic minimum to ensure that they display the same similarity behavior. Self-similarity appears locally, for positions and times close enough to $x_*$ and $t_*$, respectively. As discussed in section \ref{sec:stability}, the self-similar solution that appears prior to $t_*$ is dubbed the `closure' solution, since here the concentration remains zero $\Gamma(x_*,t)=0$ at all times. After the dimple `closes' at $t_*$, a different `leveling' solution appears, whereby the concentration starts leveling towards the final steady distribution $\Gamma(x,t)={1}$.

It is shown in section \ref{sec:stability} that the similarity exponent for dimple solutions is $\beta=2$, which can be substituted in the implicit similarity solution \eqref{eq:implicit_sol_f} to yield
\begin{equation}\label{eq:polyn_f_dimples}
    \sgn{\eta}k_\pm{f}^2+f-\eta = 0.
\end{equation}
Solutions to \eqref{eq:polyn_f_dimples} \red{represent} both (pre-singularity) closure solutions and (post-singularity) leveling solutions\red{, which have expansions around $\eta=0$ (table \ref{tab:fixed_pts}) given by $f(\eta)\sim\eta+\ii{K}\eta^2$ and $f(\eta)\sim\ii{K}-\eta$, respectively. Both of these expansions} lead to \emph{imaginary} constants $k_\pm$ when introduced in \eqref{eq:polyn_f_dimples}. \red{This means that, due to the symmetry condition $k_-=\bar{k}_+$ \eqref{eq:k_condition}, we can consider a single constant $k=k_+=-k_-=\sgn{\eta}k_\pm$.} Solving \eqref{eq:polyn_f_dimples}, 
\begin{equation}\label{eq:f_dimple_with_k}
    f(\eta) = \dfrac{-1\pm\sqrt{1+4k\eta}}{2k}.
\end{equation}

For the \emph{leveling} solution, \red{we expect $\psi(x_*,t)$, and thus $f(0)$, to be nonzero, which leads us to choose} the minus sign in \eqref{eq:f_dimple_with_k}. This sign choice must be valid for all $\eta$ since a change from $-$ to $+$ requires the square root to be zero to maintain a continuous solution, while the radicand $1+4k\eta$ can never be zero with $k$ imaginary. We also fix $\red{k=\ii}$ so that $\red{f(0)=\ii}$ (or, equivalently, $C(0)=1$). The dimple leveling solution is then
\begin{equation} \label{eq:f_dimple_leveling}
    \red{f(\eta)=\dfrac{\ii}{2}\left(\sqrt{1+4\ii\eta}+1\right)},
\end{equation}
which can alternatively be decomposed into its real and imaginary parts using the relation $\sqrt{z}=\sqrt{(|z|+\re{z})/2}+\ii\,\sgn{\im{z}}\sqrt{(|z|-\re{z})/2}$, leading to
\begin{subequations} \label{eq:f_dimple_leveling_ReIm}
    \begin{align}
        C(\eta) &= \dfrac{1}{2}\left(\sqrt{\dfrac{1}{2}\left[\sqrt{1+16\eta^2}+1\right]}+1\right),\\[4pt]
        U(\eta) &= -\dfrac{1}{2}\,\sgn{\eta}\sqrt{\dfrac{1}{2}\left[\sqrt{1+16\eta^2}-1\right]}.
    \end{align}
\end{subequations}
We note that \eqref{eq:f_dimple_leveling_ReIm} leads to $C(\eta)\sim(\sqrt{2}/2)\red{\,|\eta|^{1/2}}$ and $U(\eta)\sim-\sgn{\eta}(\sqrt{2}/2)\red{\,|\eta|^{1/2}}$ in the far field $|\eta|\to\infty$, compatible with \eqref{eq:far_field}.

For the \emph{closure} solution, the choice in \eqref{eq:f_dimple_with_k} must be the plus sign such that $f(0)=0$. \red{In the absence of an obvious choice for an integration constant $k$ (since $U(0)=C(0)=0$), we fix its value so that the far-field behavior of the closure solution is equivalent to that of the leveling solution \eqref{eq:f_dimple_leveling_ReIm}. In other words, we require $C(\eta)\sim(\sqrt{2}/2)\red{\,|\eta|^{1/2}}$ and $U(\eta)\sim\sgn{\eta}(\sqrt{2}/2)\red{\,|\eta|^{1/2}}$ as $|\eta|\to\infty$, where the sign change in $U(\eta)$ is due to the reversal in the sign of $x$ between the definitions of the similarity variable \eqref{eq:eta_fwd} and \eqref{eq:eta_bwd} for each solution. This leads to $f(\eta)\sim{e}^{\frac{\ii\pi}{4}}\eta^{1/2}$ as $\eta\to\infty$, which introduced in \eqref{eq:polyn_f_dimples} results in $k=-\ii$.} The final form of the dimple closure solution is then
\begin{equation} \label{eq:f_dimple_closure}
    \red{f(\eta)=\dfrac{\ii}{2}\left(\sqrt{1-4\ii\eta}-1\right)},
\end{equation}
from which we obtain
\begin{subequations} \label{eq:f_dimple_closure_ReIm}
    \begin{align}
        C(\eta) &= \dfrac{1}{2}\left(\sqrt{\dfrac{1}{2}\left[\sqrt{1+16\eta^2}+1\right]}-1\right),\\[4pt]
        U(\eta) &= \dfrac{1}{2}\,\sgn{\eta}\sqrt{\dfrac{1}{2}\left[\sqrt{1+16\eta^2}-1\right]}.
    \end{align}
\end{subequations}
The choice of \eqref{eq:f_dimple_leveling} and \eqref{eq:f_dimple_closure} having an equivalent far-field behavior ensures that the multiplicative constant $A$ in the similarity formulation \eqref{eq:ansatz} is the same for both the closure and the leveling solutions. This fact simplifies calculations, since it then suffices to \red{compute} $A$ for only one of the two solutions.

The middle column of figure \ref{fig:dimple} shows how the exact solutions converge to the closure self-similar profiles given by \eqref{eq:f_dimple_closure} before the singularity, for the three distinct initial profiles $\Gamma_0(x)$ considered. Likewise, the right-most column in figure \ref{fig:dimple} illustrates that exact solutions converge to the leveling solution \eqref{eq:f_dimple_leveling} after the singularity. Since the similarity solutions are only valid locally, the agreement between the rescaled profiles is always improved as $t\to{t_*}$ or as $x\to{x_*}$.

\subsection{Hole solutions} \label{sec:filling_sols_hole}
\begin{figure}
    \centering 
    \subfloat{\includegraphics[]{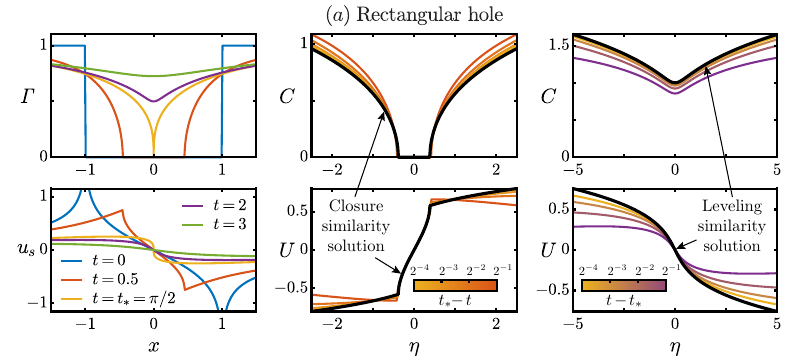}\label{fig:hole_rectangular}} \\[-2pt]
    \subfloat{\includegraphics[]{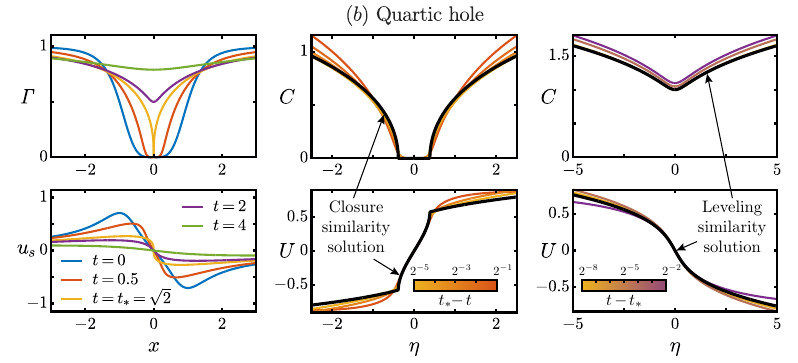}\label{fig:hole_quartic}} \\[-2pt]
    \subfloat{\includegraphics[]{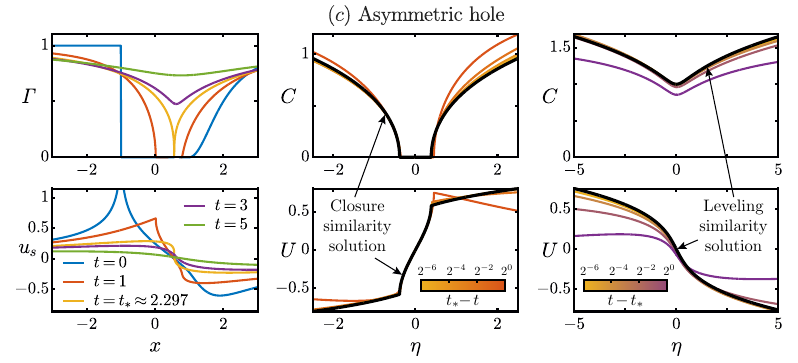}\label{fig:hole_asymmetric}}
    \caption{Filling solutions with $Pe_s^{-1}=0$, for initial `hole' distributions of surfactant given by $(a)$ a rectangular hole, $(b)$ a quartic hole, and $(c)$ an asymmetric hole, with their functional forms given in appendix \ref{sec:apdx_hilbert_transforms}. For each example, the left column shows the concentration $\Gamma(x,t)$ and interfacial velocity $u_s(x,t)$ obtained through the exact solution of \eqref{eq:Hopf}. The middle and right columns show exact solutions rescaled in similarity variables (color curves), superimposed to the closure \eqref{eq:f_hole_closure_ReIm} and leveling \eqref{eq:f_hole_leveling_ReIm} similarity solutions (black curves) valid prior and subsequent to the singularity, respectively. The asymmetric hole has $x_*\approx{0.562}$, otherwise $x_*=0$.}
    \label{fig:hole}
\end{figure}

Contrary to dimples, hole similarity solutions (with $\beta=3/2$) had not yet been identified for the complex inviscid Burgers equation \eqref{eq:Hopf_pde}. One of the simplest examples of a distribution of surfactant that satisfies this condition is a `rectangular hole' with $\Gamma_0(x)=H(|x|-1)$, where $H(x)$ is the Heaviside step function, which we illustrate in figure \ref{fig:hole_rectangular}. 
For this case, the initial surfactant profile is even and therefore its Hilbert transform $u_{s0}(x)=\hil{\Gamma_0(x)}$ is odd, which implies (see appendix \ref{sec:apdx_closure_time}) that the position of the singularity is $x_*=0$, its velocity $u_*=0$, and the closure time $t_*=\pi/2$. \red{One can also} easily verify that $\Gamma_0''(0)=u_{s0}''(0)=0$ and thus the condition for hole self-similar solutions $\psi_0''(x_*-u_*t_*)=\psi_0''(0)=0$ is satisfied. Figure \ref{fig:hole_rectangular} depicts the exact evolution of the rectangular hole according to the inviscid Burgers equation \eqref{eq:Hopf_pde}. Like for the case of dimples, the distribution first goes through a `closure' phase where surfactant is advected inwards but remains $\Gamma(0,t)=0$ \red{at the origin}. However, figure \ref{fig:hole_rectangular} \red{also shows} that the self-similar dynamics before the closure time $t_*$ must be different from those of the dimple, since the solution retains a finite interval around $x_*=0$ where the concentration remains zero. After $t_*=\pi/2$, the concentration at the origin starts `leveling', with $\Gamma(0,t)>0$, until the profile reaches a homogeneous distribution $\Gamma(x,t)=1$ as $t\to\infty$.

In this case, the self-similar solutions can also be obtained in closed form by substituting $\beta=3/2$ in the implicit solution given by \eqref{eq:implicit_sol_f}, which leads to
\begin{equation} \label{eq:polyn_f_holes}
    k_\pm f^3 + f - \eta = 0.
\end{equation}
\red{Substituting the expansions around $\eta=0$ found in table \ref{tab:fixed_pts} for the closure ($f\sim\eta+K\eta^3$) and leveling solutions ($f\sim-\ii{K}-\eta/2$) into \eqref{eq:polyn_f_holes} leads to \emph{real} constants $k_\pm$. Combined with the symmetry condition \eqref{eq:k_condition}, this indicates that} for either solution we can consider a single constant $k=k_-=k_+$, as in the case of spreading pulse solutions of section \ref{sec:spreading_sols_no_diff}.

We first solve \eqref{eq:polyn_f_holes} for the case of the (post-singularity) leveling solution. We again choose to fix $\red{f(0)=\ii}$ or, equivalently, $C(0)=1$, which results in $k=1$. The discriminant of the cubic \eqref{eq:polyn_f_holes} is then $-4-27\eta^2$, which is negative for any $\eta$. This implies that \eqref{eq:polyn_f_holes}, for any given value of $\eta$, has one real and two complex conjugate solutions that can be obtained using standard methods for solving cubic equations \citep{Cox2012-kk}. Choosing the complex solution with a \red{positive} imaginary part results in the hole leveling solution
\begin{equation} \label{eq:f_hole_leveling}
    \red{f(\eta) = e^{\frac{\ii\pi}{3}}\sqrt[3]{\dfrac{1}{2}\left[\sqrt{\eta^2+\dfrac{4}{27}}-\eta\right]} + e^{\frac{2\ii\pi}{3}}\sqrt[3]{\dfrac{1}{2}\left[\sqrt{\eta^2+\dfrac{4}{27}}+\eta\right]}}.
\end{equation}
Furthermore, since the arguments of the two cubic roots in \eqref{eq:f_hole_leveling} are always real and positive, it is straightforward to decompose the expression into
\begin{subequations} \label{eq:f_hole_leveling_ReIm}
    \begin{alignat}{3}
        C(\eta) &= &\,\,\dfrac{\sqrt{3}}{2}&\left(\sqrt[3]{\dfrac{1}{2}\left[ \sqrt{\eta^2+\dfrac{4}{27}}-\eta\right]}+\sqrt[3]{\dfrac{1}{2}\left[\sqrt{\eta^2+\dfrac{4}{27}}+\eta\right]}\,\right),\\[4pt]
        U(\eta) &= &\dfrac{1}{2}&\left(\sqrt[3]{\dfrac{1}{2}\left[ \sqrt{\eta^2+\dfrac{4}{27}}-\eta\right]}-\sqrt[3]{\dfrac{1}{2}\left[\sqrt{\eta^2+\dfrac{4}{27}}+\eta\right]}\,\right),
    \end{alignat}
\end{subequations}
where the far-field behavior $|\eta|\to\infty$ is in this case given by $C(\eta)\sim(\sqrt{3}/2)\red{\,|\eta|^{1/3}}$ and $U(\eta)\sim-\,\sgn{\eta}(1/2)\red{\,|\eta|^{1/3}}$.

In the case of the closure solution, we \red{again} choose the integration constant such \red{that its far field} is equivalent to that of the leveling solution, which requires $k=-1$. As in the case of dimples, this choice of far field ensures that the scaling constant $A$ in the similarity formulation \eqref{eq:ansatz} is the same for both closure and leveling. The discriminant of \eqref{eq:polyn_f_holes} is then $4-27\eta^2$, which indicates that its three solutions are real for $|\eta|\leq\sqrt{4/27}=2\sqrt{3}/9$, whereas for $|\eta|\geq2\sqrt{3}/9$ there is one real and two complex conjugate solutions. The only way to ensure a continuous solution with $C(\eta)>0$ is to define $f(\eta)$ piecewise, with one of the three solutions of the cubic valid for $\red{\eta\leq-2\sqrt{3}/9}$, and another one valid for $\red{\eta\geq-2\sqrt{3}/9}$. Such a solution is
\begin{equation}
    \red{f(\eta)=}
    \begin{cases}
            \red{e^\frac{\ii\pi}{3}\left(\sqrt[3]{-\dfrac{1}{2}\left[\sqrt{\eta^2-\dfrac{4}{27}}-\eta\right]}-\sqrt[3]{-\dfrac{1}{2}\left[\sqrt{\eta^2-\dfrac{4}{27}}+\eta\right]}\,\,\right)} & \red{\text{if }\eta\leq-\dfrac{2\sqrt{3}}{9}},\\[20pt]
            \red{e^\frac{\ii\pi}{3}\left(\sqrt[3]{\phantom{-}\dfrac{1}{2}\left[\sqrt{\eta^2-\dfrac{4}{27}}+\eta\right]}-\sqrt[3]{\phantom{-}\dfrac{1}{2}\left[\sqrt{\eta^2-\dfrac{4}{27}}-\eta\right]}\,\,\right)} & \red{\text{if }\eta\geq-\dfrac{2\sqrt{3}}{9}},
    \end{cases}
\end{equation}
which can be expressed more compactly \red{as}
\begin{equation}\label{eq:f_hole_closure}
    \red{f(\eta) = e^\frac{\ii\pi}{3} \left( \sqrt[3]{\frac{1}{2}\left[\sqrt{\eta+\frac{2\sqrt{3}}{9}}\sqrt{\eta-\frac{2\sqrt{3}}{9}}+\eta\right]} - \sqrt[3]{\dfrac{1}{2}\left[\sqrt{\eta+\dfrac{2\sqrt{3}}{9}}\sqrt{\eta-\dfrac{2\sqrt{3}}{9}}-\eta\right]} \,\right)}.
\end{equation}
For $|\eta|\geq2\sqrt{3}/9$, the argument of the cubic roots in \eqref{eq:f_hole_closure} is always real, leading to a straightforward decomposition into real and imaginary parts. However, in the case of $|\eta|\leq2\sqrt{3}/9$, the real and imaginary parts of the solution can only be obtained through the trigonometric solution of the cubic \citep{Cox2012-kk}. In summary, the final form of the similarity solutions $C(\eta)$ and $U(\eta)$ is 
\begin{subequations} \label{eq:f_hole_closure_ReIm}
    \begin{align}
        C(\eta) &= 
        \begin{cases}
            0 & \text{for }|\eta|\leq\dfrac{2\sqrt{3}}{9},\\[15pt]
            \dfrac{\sqrt{3}}{2}\left(\sqrt[3]{\dfrac{1}{2}\left[|\eta|+\sqrt{\eta^2-\dfrac{4}{27}}\right]}-\sqrt[3]{\dfrac{1}{2}\left[|\eta|-\sqrt{\eta^2-\dfrac{4}{27}}\right]}\,\right) &\text{for }|\eta|\geq\dfrac{2\sqrt{3}}{9},
        \end{cases}\\[4pt]
        U(\eta) &= 
        \begin{cases}
            \dfrac{2\sqrt{3}}{3}\,\sin\left[\dfrac{1}{3}\arcsin\left(\dfrac{3\sqrt{3}}{2}\,\eta\right)\right] & \text{for }|\eta|\leq\dfrac{2\sqrt{3}}{9},\\[15pt]
            \dfrac{1}{2}\,\sgn{\eta}\left(\sqrt[3]{\dfrac{1}{2}\left[|\eta|+\sqrt{\eta^2-\dfrac{4}{27}}\right]}\hspace{-1.5pt}+\hspace{-1.5pt}\sqrt[3]{\dfrac{1}{2}\left[|\eta|-\sqrt{\eta^2-\dfrac{4}{27}}\right]}\,\right) &\text{for }|\eta|\geq\dfrac{2\sqrt{3}}{9},
        \end{cases}
    \end{align}
\end{subequations}
where we can \red{verify} that its far field \red{as $|\eta|\to\infty$}, which is given by $C(\eta)\sim(\sqrt{3}/2)\red{\,|\eta|^{1/3}}$ and $U(\eta)\sim\sgn{\eta}(1/2)\red{\,|\eta|^{1/3}}$, is equivalent to that of the leveling solution.

The center and right columns of figure \ref{fig:hole_rectangular} show that the exact solution, when appropriately rescaled using \eqref{eq:ansatz}, converges to the closure solution \eqref{eq:f_hole_closure_ReIm} before $t_*$ and to the leveling solution \eqref{eq:f_hole_leveling_ReIm} after $t_*$. Other initial surfactant profiles also lead to these similarity solutions, as long as the condition $\psi_0''(x_*-t_*u_*)=0$ is met. It is worth noting that the initial surfactant distribution does not need to be zero at a finite interval to tend to a self-similar solution for a hole, as exemplified by the `quartic hole' initial condition $\Gamma_0(x)=x^4/(1+x^4)$, whose evolution is displayed in figure \ref{fig:hole_quartic}. This initial profile also has even symmetry, leading to $x_*=0$, $u_*=0$, and $t_*=\sqrt{2}$ (see appendix \ref{sec:apdx_closure_time}), but its initial concentration is zero only at the origin $x_*=0$. Regardless, since $\psi_0''(x_*-u_*t_*)=\psi_0''(0)=0$, the \red{profiles evolve towards} the closure solution \eqref{eq:f_hole_closure} for $t<t_*$ (center column in figure \ref{fig:hole_quartic}) and \red{towards the leveling solution} \eqref{eq:f_hole_leveling} for $t>t_*$ (right column). Even though $\Gamma_0(x)$ is zero at a single point, the concentration `flattens' as $t\to{t_*}$ to converge towards a solution that is zero at a finite interval. 

The last example illustrated in figure \ref{fig:hole_asymmetric} is an \emph{asymmetric} initial condition. For this case, we have that $x_*\neq{0}$ and $u_*\neq{0}$, although their values can still be calculated from the method of characteristics, as detailed in appendix \ref{sec:apdx_closure_time}. The self-similar dynamics are still governed by the solutions \eqref{eq:f_hole_closure} and \eqref{eq:f_hole_leveling} although, since the point of the singularity moves with $u_*\neq{0}$, the \red{similarity} variable must \red{account} for a frame of reference moving with the \red{singular point}. This leads to a more general similarity ansatz
\begin{subequations} \label{eq:ansatz_moving}
    \begin{align}
        \psi(x,t) &= u_* + A|t-t_*|^{\beta-1}f(\eta), \\
        \eta &= \sgn{t-t_*} \dfrac{x-\left[x_*+u_*(t-t_*)\right]}{A|t-t_*|^\beta}.  
    \end{align}
\end{subequations}
Equation \eqref{eq:ansatz_moving} accounts for the moving \red{singular point}, whose position is $x_s(t)=x_*+u_*(t-t_*)$. Note that $x_s(t_*)=x_*$, whereas $x_s(0)=x_*-u_*t_*$, which is the departure \red{position} of the moving \red{singular point}.

\section{Conclusions} \label{sec:conclusions}
Quantitatively describing Marangoni flows induced by surfactant is a central problem in interfacial fluid dynamics, due to their prevalence in environmentally and industrially relevant multiphase flows. Motivated by recent theoretical progress, we have investigated the two-dimensional spreading problem for a deep, viscous subphase in terms of its self-similarity. The analysis reveals a \red{rich} structure with six distinct similarity solutions and three different exponents $\beta$, \red{listed} in table \ref{tab:summary},
\begin{table}
\begin{center}
\def~{\hphantom{0}}
\begin{tabular}{c c c c c c c}
     Name & $Pe_s^{-1}$ & Kind & Validity & Exponent $\beta$ & Similarity variable $\eta$ & Solution $f(\eta)$ \\\hline
     Pulse  & 0 & First & $t\gg{1}$ & 1/2 & $\dfrac{x-x_*}{\left(\dfrac{M_0}{\pi}\,t\right)^{1/2}}$ & \eqref{eq:f_pulses} \\[25pt]
     Pulse & $>0$ & First & $t\gg{1}$ & 1/2 & $\dfrac{x-x_*}{\left(\dfrac{2}{Pe_s}\,t\right)^{1/2}}$ & \eqref{eq:f_pulses_diff} \\[25pt]
     Dimple closure & 0 & Second & $t\lesssim{t_*}$ & 2 & $\dfrac{[x_*+u_*(t-t_*)]-x}{A(t_*-t)^2}$ & \eqref{eq:f_dimple_closure} \\[15pt]
     Dimple leveling & 0 & Second & $t\gtrsim{t_*}$ & 2 & $\dfrac{x-[x_*+u_*(t-t_*)]}{A(t-t_*)^2}$ & \eqref{eq:f_dimple_leveling} \\[15pt]
     Hole closure & 0 & Second & $t\lesssim{t_*}$ & 3/2 & $\dfrac{[x_*+u_*(t-t_*)]-x}{A(t_*-t)^{3/2}}$ & \eqref{eq:f_hole_closure} \\[15pt]
     Hole leveling & 0 & Second & $t\gtrsim{t_*}$ & 3/2 & $\dfrac{x-[x_*+u_*(t-t_*)]}{A(t-t_*)^{3/2}}$ & \eqref{eq:f_hole_leveling}
\end{tabular}
\caption{Summary of the six similarity solutions found in this study, indicating the equation number of each solution $f(\eta)$ obtained in closed form. Here, $M_0$ is the dimensionless surfactant mass as defined in \eqref{eq:conservation_mass}, and $Pe_s$ is the P\'eclet number given by \eqref{eq:peclet}. In the case of pulses, the reference position $x_*$ is the center of mass of the surfactant distribution given by \eqref{eq:x_star_pulses}, and in the case of dimples and holes $x_*$ is the `closure position' at which the solution has a weak singularity, which can be calculated a priori from the initial conditions as described in appendix \ref{sec:apdx_closure_time}. The parameters $t_*$ and $u_*$ are the closure time and instantaneous velocity of the \red{singular point}, respectively, and can also be calculated using appendix \ref{sec:apdx_closure_time}. For solutions of the second kind, the constant $A$ depends on local properties of the initial condition $\Gamma_0(x)$.}
\label{tab:summary}
\end{center}
\end{table}
all of which can be obtained in closed \red{form}.

In section \ref{sec:spreading_sols}, we derive one similarity solution without diffusion ($Pe_s^{-1}=0$) and another with diffusion ($Pe_s^{-1}>0$) for the case of pulses of surfactant, both of which are valid at long times $t\gg{1}$. \red{The solution with $Pe_s^{-1}=0$ is in fact only valid for times up to $1\ll{t}\ll{Pe_s}$, since at $t=O(Pe_s)$ the interfacial velocity is expected to become small enough for diffusion to be comparable to advection}. These two \red{spreading} solutions are equivalent to the ones previously identified by \cite{Thess1996-bc} and \cite{Bickel2022-sa}, respectively, through different methods. In addition to their derivation, we have also shown (appendix \ref{sec:apdx_invariants}) how to calculate the center of mass $x_*$ around which these solutions appear, something particularly useful when the initial surfactant distribution is asymmetric or the combination of several pulses. Since their similarity exponent is $\beta=1/2$, these pulse solutions are analogous to a diffusive process where the surfactant peak decreases as $\Gamma\propto{t}^{-1/2}$, and its front spreads as $x_f\propto{t}^{1/2}$. These two solutions can therefore be used to obtain \emph{effective} surfactant diffusivities resulting from the Marangoni flow, as detailed by \cite{Bickel2022-sa}. We also note that the solutions $N\to{O}$ in the phase plane (figure \ref{fig:phase_plane}) that have $1/2<\beta<1$ are also spreading and, in principle, physically admissible in terms of their stability (section \ref{sec:stability}). Therefore, we postulate that surfactant pulses that decay too slowly in the far field to have a well-defined mass $M_0$ might display this kind of self-similar solution.

Section \ref{sec:filling_sols} is concerned with surfactant distributions that are locally depleted and flow inwards, for which similarity only occurs for $Pe_s^{-1}=0$. We have provided the first derivation of two similarity solutions with $\beta=2$ \red{and another} two with $\beta=3/2$. Through insights provided by stability analysis (section \ref{sec:stability}) and the complex method of characteristics, we have also provided a quantitative criterion to determine if a given initial surfactant profile will develop similarity with $\beta=2$, in which case we call such profile a `dimple', or with $\beta=3/2$, in which case we call it a `hole'. Aside from providing valuable information about the spatial and temporal structure of the evolution of surfactant, these solutions also allow to calculate effective local properties of the flow. For example, from the similarity ansatz \eqref{eq:ansatz_psi}, we can deduce that the concentration at the centerline $x_*$ of an interfacial strip that is depleted of surfactant is
\begin{subequations}
    \begin{align}
        \Gamma(x_*,t) = 
        \begin{cases}
            0 & \text{if }0\leq{t}\leq{t}_*,\\
            A(t-t_*) & \text{if }{t}\gtrsim{t}_*,
        \end{cases}
    \end{align}
\end{subequations}
for dimples, while for holes
\begin{subequations}
    \begin{align}
        \Gamma(x_*,t) = 
        \begin{cases}
            0 & \text{if }0\leq{t}\leq{t}_*,\\
            A(t-t_*)^{1/2} & \text{if }{t}\gtrsim{t}_*,
        \end{cases}
    \end{align}
\end{subequations}
where $t_*$ can be obtained exactly if \red{$\Gamma_0(x)$} is known, as detailed in appendix \ref{sec:apdx_closure_time}.

Since local surfactant concentrations are challenging to measure experimentally, one can also derive expressions for the centerline interfacial shear, which reads
\begin{subequations}
    \begin{align}
        \pd{u_s}{x}{}(x_*,t) = 
        \begin{cases}
            \dfrac{1}{t_*-t} & \text{if }{t}\lesssim{t}_*,\\[10pt]
            \dfrac{1}{t-t_*} & \text{if }{t}\gtrsim{t}_*,
        \end{cases}
    \end{align}
\end{subequations}
for dimples, and 
\begin{subequations}
    \begin{align}
        \pd{u_s}{x}{}(x_*,t) = 
        \begin{cases}
            \dfrac{1}{t_*-t} & \text{if }{t}\lesssim{t}_*,\\[10pt]
            \dfrac{1}{2(t-t_*)} & \text{if }{t}\gtrsim{t}_*,
        \end{cases}
    \end{align}
\end{subequations}
for holes. These expressions for the interfacial shear are in principle obtainable by measuring the interfacial velocity field in experiments, and should be valid for times sufficiently near $t_*$, but not too close to the singularity for surface diffusion to locally regularize the interfacial velocity field. The expressions do not depend on any scaling constant $A$, and the only parameter involved, $t_*$, can be either calculated exactly if $\Gamma_0(x)$ is known, or measured from experimental data. 

This taxonomy of self-similar solutions provides insights into the behavior of Marangoni flows on a deep fluid subphase, in the limit of low Reynolds and capillary numbers. A natural question arises from this analysis: given an arbitrary initial distribution of surfactant, will it always evolve to one of these similarity solutions? We expect that any profile decaying in the far field will eventually converge to `spreading' similarity solutions, either with diffusion \eqref{eq:f_pulses_diff} or without it \eqref{eq:f_pulses} if $Pe_s\gg{1}$. This is consistent with general self-similar behavior appearing in scale-free physical systems at long times \citep{Barenblatt1996-hd}, and we have observed it even with multiple surfactant pulses (see Figure \ref{fig:spreading_no_diff_double}). On the other hand, `filling' self-similar solutions appear locally for depleted distributions of surfactant, but only in the absence of diffusion and if the initial concentration $\Gamma_0(x)$ is exactly zero at some point. We have conducted a preliminary comparison between the `hole' and `dimple' solutions and simulations in more realistic scenarios, which include small amounts of background endogenous surfactant \citep[see][]{Grotberg1995-ph} and a finite diffusion \citep[as also analyzed by][]{Crowdy2021-vy}. We found that surface diffusion, no matter how small, locally regularizes the singularities in the derivatives, but the similarity solutions still provide a good approximation of the dynamics at high $Pe_s\gg{1}$. This suggests that any profile that does not decay in the far field but is locally depleted could potentially be approximated by self-similar solutions, as long as the minimum value of $\Gamma_0(x)$ and diffusion are both sufficiently low. A detailed analysis, which could perhaps be achieved perturbatively, could provide further insights into the generality of self-similar behavior given arbitrary initial conditions. Similarly, it is worth asking if a self-similarity approach would yield similar insights in an axisymmetric geometry, since this work deals exclusively with a planar, two-dimensional domain. \red{The axisymmetric problem has} a more complicated nonlocal closure \citep{Bickel2022-sa} for which it appears that no reformulations like Burgers equation exist, but the tools of self-similarity could still be applied for nonlocal problems \citep[as in][for example]{Lister1989-so}.

\section*{Acknowledgements}
F.T-C. acknowledges support from a Distinguished Postdoctoral Fellowship from the Andlinger Center for Energy and the Environment. We thank the National Science Foundation for partial support through grant CBET 2127563.

\section*{Declaration of interests}
The authors report no conflict of interest.

\appendix
\section{Construction of the phase plane} \label{sec:apdx_phase_plane}
We first recast the autonomous ODE \eqref{eq:ode_g}, which governs the behavior of the complex similarity solution $f(\eta)=\eta\,{g}(\eta)$ in the limit $Pe_s^{-1}=0$, as
\begin{equation}
    \td{g}{\ln|\eta|}{} = \dfrac{g(1-g)(\bar{g}-\beta)}{|g-\beta|^2},
\end{equation}
with the overbar indicating complex conjugation. Since the right-hand side of the above ODE has a singularity at $g=\beta$, we reparametrize the equation \citep[as in, for instance,][]{Slim2004-fo} in terms of an auxiliary variable $\chi$, leading to
\begin{subequations} \label{eq:ode_g_chi}
\begin{align}
    \td{g}{\chi}{} &= g(1-g)(\bar{g}-\beta), \label{eq:ode_g_chi_g}\\[2pt]
    \td{\ln|\eta|}{\chi}{} &= |g-\beta|^2. \label{eq:ode_g_chi_ln_eta}
\end{align}
\end{subequations}
Since, by virtue of equation \eqref{eq:ode_g_chi_ln_eta} above, we have that $\dd\ln|\eta|/\dd\chi \geq 0$, then integrating the system in terms of $\chi$ instead of $\ln|\eta|$ does not change the direction of trajectories in the phase space $(\re{g},\im{g})$, unlike in other more complicated systems of equations such as the one considered in \cite{Slim2004-fo}. The three fixed points of equation \eqref{eq:ode_g_chi_g} are given by $g=0$, $g=1$ and $g=\beta$, and linearization around each of them \citep[][]{Strogatz2018-tt} reveals their type, as well as the asymptotic form of the solution around each of them (points $O$, $P$ and $S$ in table \ref{tab:fixed_pts}).

We integrate \eqref{eq:ode_g_chi} numerically using the built-in MATLAB integrator \texttt{ode15s}. The initial condition of \eqref{eq:ode_g_chi_ln_eta} is chosen as $\left.(\ln|\eta|)\right|_{\chi=0}=-K$, with $K\gg{1}$ to represent a point close to the origin $\eta\approx{0}$. The initial values of $g$ are seeded close to the fixed points of the system such that $g(\chi=0)=g_0+{\Delta}e^{i\theta}$, with $\Delta\ll{1}$ and $\theta$ real constants, and where $g_0$ is the value of $g$ at each fixed point. We integrate \eqref{eq:ode_g_chi_g} forward in $\chi$ if $g(\chi=0)$ lies on an unstable direction around the fixed point, and backward in $\chi$ if $g(\chi=0)$ lies on a stable direction. Integration proceeds until $\ln|\eta|$ reaches a target value $\ln|\eta|=K\gg{1}$, denoting the far field $|\eta|\to\infty$. The resulting trajectories are shown in figure \ref{fig:phase_plane}.

We also consider the behavior of trajectories as $|g|\to\infty$, which can be illustrated by studying the fixed points of the dynamical system given by the reciprocals $1/\re{g}=\eta/U$ and $\red{1/\im{g}=\eta/C}$. Splitting the complex ODE \eqref{eq:ode_g} into its real and imaginary parts, and changing variables $\tu\defeq{1}/\re{g}$ and $\red{\tc\defeq1/\im{g}}$, we obtain the system of ODEs:
\begin{subequations} \label{eq:ode_g_reciprocals}
\begin{align}
    \td{\tu}{\ln|\eta|}{} &= \tu\,\dfrac{\left[\tu^2(1+(\beta-1)\tu)-\tc^2(\tu-1)(1-\beta\tu)\right] }{\tu^2 + \tc^2(1-\beta\tu)^2}, \label{eq:ode_g_reciprocals_u} \\[2pt]
    \td{\tc}{\ln|\eta|}{} &= \tc\,\dfrac{ \left[\tu^2+\tc^2(1-2\beta\tu+\beta\tu^2)\right] }{\tu^2 + \tc^2(1-\beta\tu)^2} \label{eq:ode_g_reciprocals_c}.
\end{align}
\end{subequations}
The fixed points of \eqref{eq:ode_g_reciprocals} are $(\tu,\tc)=(0,0)$, which represents $\red{(\re{g},\im{g})\to(\pm\infty,\pm\infty)}$, and $(\tu,\tc)=((1-\beta)^{-1},0)$, which represents $\red{(\re{g},\im{g})\to(1-\beta,\pm\infty)}$. Linearization around these two points leads to the rows of table \ref{tab:fixed_pts} corresponding to points $N$ and $R$.

Finally, the behavior of solutions for $\re{g}\to\pm\infty$ and $\red{\im{g}=0}$ can only be determined by examining the dynamical system given by the reciprocal $1/\re{g}=\eta/U$\red{, retaining} the imaginary part $\red{\im{g}}$. Changing variables $\tu\defeq{1}/\re{g}$ and $\red{\hc\defeq\im{g}}$, we obtain a dynamical system given by
\begin{subequations} \label{eq:ode_g_reciprocal_re}
\begin{align}
    \td{\tu}{\ln|\eta|}{} &= \tu\,\dfrac{\left[\tu\hc^2(1+(\beta-1)\tu)-(\tu-1)(1-\beta\tu)\right]}{(1-\beta\tu)^2+\tu^2\hc^2}, \label{eq:ode_g_reciprocal_re_u} \\[2pt]
    \td{\hc}{\ln|\eta|}{} &= \hc\,\dfrac{\left[1-2\beta\tu+\beta\tu^2+\tu^2\hc^2\right]}{(1-\beta\tu)^2+\tu^2\hc^2}. \label{eq:ode_g_reciprocal_re_c}
\end{align}
\end{subequations}
The only fixed point of \eqref{eq:ode_g_reciprocal_re} is $(\tu,\hc)=(0,0)$, which represents $\red{(\re{g},\im{g})\to(\pm\infty,0)}$. Linearization of \eqref{eq:ode_g_reciprocal_re} around \red{it} results in the row of table \ref{tab:fixed_pts} corresponding to point $M$.

\section{Interpretation of the phase plane} \label{sec:apdx_interpret_plane}
In order to interpret the phase plane in figure \ref{fig:phase_plane}, it is useful to note two facts about the sign of solutions. First, from the self-similar ansatz \eqref{eq:Gamma_f} and the facts that $\alpha=\beta-1$ \red{and $A=B$}, we have at the origin $x=x_*$ that $\Gamma(x_*,t)=A|t-t_*|^{\beta-1}C(0)$, illustrating that, if $C(0)>{0}$, values of $0<\beta<1$ will result in surfactant locally decreasing in time (i.e., spreading solutions), whereas exponents $\beta>1$ represent locally increasing surfactant (i.e.\red{,} filling solutions). Consequently, solutions with $0<\beta<1$ must lead to a (locally) outward flow as in figure \ref{fig:problem_setup_spreading}, with $u_s$ positive for $x>x_*$ and $u_s$ negative for $x<x_*$ or, in other words, $(x-x_*)u_s>0$. On the other hand, solutions with $\beta>1$ must lead to $(x-x_*)u_s<0$ locally around the origin as in figure \ref{fig:problem_setup_filling}.

\begin{figure}
    \centering 
    \includegraphics[]{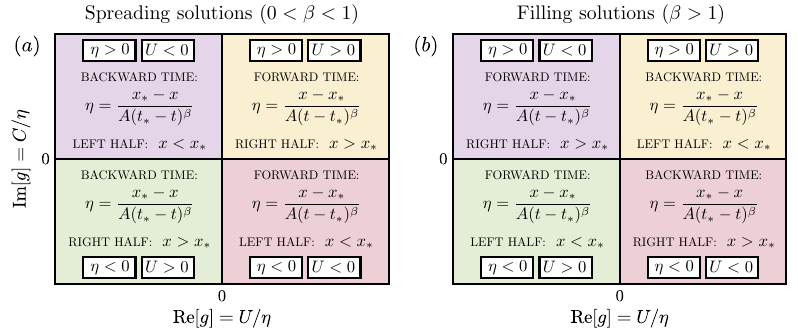}
    \caption{Physical interpretation of each quadrant of the phase plane in figure \ref{fig:phase_plane}, for $(a)$ spreading solutions with an outward flow (as in figure \ref{fig:problem_setup_spreading}), and $(b)$ filling solutions with an inward flow (as in figure \ref{fig:problem_setup_filling}). In order for the concentration $\Gamma(x,t)$ to be strictly non-negative, each quadrant must correspond to a specific definition of the similarity variable, either forward-time as in equation \eqref{eq:eta_fwd} or backward-time as in equation \eqref{eq:eta_bwd}. In addition, each quadrant represents one half of the real line, either $x>x_*$ or $x<x_*$.}
    \label{fig:phase_plane_interpret}
\end{figure}

Second, physical solutions require $\Gamma(x,t)\geq{0}$ and therefore also $C(\eta)\geq{0}$. Since each quadrant of the phase plane has a fixed sign of $C(\eta)/\eta$ and $U(\eta)/\eta$, it then follows that each quadrant must also have a fixed sign of $\eta$ and $U(\eta)$ individually. These sign restrictions lead to a unique meaning for each quadrant of the phase plane, as illustrated in figure \ref{fig:phase_plane_interpret}.
For a given value of $\beta$, each quadrant must represent either a forward-time \eqref{eq:eta_fwd} or backward-time \eqref{eq:eta_bwd} scaling, as well as necessarily belong to either the right half of the real line (i.e., $x>x_*$), or to the left half (i.e., $x<x_*$).

\section{Invariants of the problem} \label{sec:apdx_invariants}
Direct integration of the surfactant conservation law given by equation \eqref{eq:surfactant_pde} yields
\begin{equation}
    \td{}{t}{}\int_{-\infty}^{\infty}\Gamma\dd{x} + \left.\left(u_s\Gamma\right)\right|_{-\infty}^{\infty} = \dfrac{1}{Pe_s}\left.\pd{\Gamma}{x}{}\right|_{-\infty}^{\infty}
\end{equation}
and, since $u_s(x,t)$ necessarily decays and $\Gamma(x,t)$ can be at most constant in the far field,
\begin{equation}
    \td{}{t}{}\int_{-\infty}^{\infty}\Gamma\dd{x} = 0.
    \label{eq:conservation_mass}
\end{equation}
Equation \eqref{eq:conservation_mass} implies that the total mass $M_0$ of surfactant, as defined in \eqref{eq:def_M0}, is conserved in time. This holds as long as the integral given by \eqref{eq:def_M0} exists, which is the case for initial pulses of surfactant with $\Gamma_0(x)$ decaying sufficiently quickly as $|x|\to\infty$. 

Furthermore, multiplying \eqref{eq:surfactant_pde} by $x$ and applying the chain rule, we obtain
\begin{equation}
    \pd{\left(x\Gamma\right)}{t}{} + \pd{}{x}{}\left(xu_s\Gamma\right) - u_s\Gamma = \dfrac{1}{Pe_s}\left[\pd{}{x}{}\left(x\pd{\Gamma}{x}{}\right)-\pd{\Gamma}{x}{}\right],
\end{equation}
which, upon integration, yields
\begin{equation}
    \td{}{t}{}\int_{-\infty}^{\infty}x\Gamma\dd{x} + \left.\left(xu_s\Gamma\right)\right|_{-\infty}^{\infty} - \int_{-\infty}^{\infty}u_s\Gamma\dd{x} = \dfrac{1}{Pe_s}\left[\left.\left(x\pd{\Gamma}{x}{}\right)\right|_{-\infty}^{\infty}-\left.\Gamma\right|_{-\infty}^{\infty}\right].
    \label{eq:integrated_pde_times_x}
\end{equation}
Since the far-field concentration of surfactant can be at most constant with the same values as $x\to\infty$ and as $x\to-\infty$, all the far-field flux terms in \eqref{eq:integrated_pde_times_x} vanish as long as the product $u_s\Gamma$ decays at least as $u_s\Gamma\sim{x}^{-1}$ as $|x|\to\infty$. Remarkably, since in this problem $u_s=\hil{\Gamma}$, the integral term in \eqref{eq:integrated_pde_times_x} also vanishes due to the orthogonality condition of the Hilbert transform \citep[][]{King2009-hr}, namely
\begin{equation}
    \int_{-\infty}^{\infty}u_s\Gamma\dd{x} = \int_{-\infty}^{\infty}\hil{\Gamma}\Gamma\dd{x} = 0.
\end{equation}
All the above implies that the first moment $M_1$ of the surfactant distribution, as defined in \eqref{eq:def_M01}, is also conserved, satisfying
\begin{equation}
    \td{}{t}{}\int_{-\infty}^{\infty}x\Gamma\dd{x} = 0
\end{equation}
as long as $x\Gamma_0(x)$ decays sufficiently quickly for the above integral to exist. 

\section{Closure time of dimple/hole distributions} \label{sec:apdx_closure_time}
The solution of the inviscid Burgers problem \eqref{eq:Hopf} can be written implicitly using the method of characteristics \citep{Crowdy2021-vy}, yielding
\begin{equation}\label{eq:sol_method_char}
    \psi(x,t) = \psi_0(x-t\,\psi(x,t)).
\end{equation}
Defining the characteristic variable $\xi(x,t)\defeq x-t\,\psi(x,t)$ and differentiating \eqref{eq:sol_method_char} yields
\begin{align} \label{eq:dx_char}
    \pd{\psi}{x}{}(x,t) = \dfrac{ \psi_0'(\xi) }{ 1 + t\,\psi_0'(\xi) }.
\end{align}
Therefore, singularities in the solution derivatives occur when $1+t_*\,\psi_0'(\xi_*)=0$, for some characteristic $\xi_*=x_*-t_*\psi(x_*,t_*)$ crossing the singularity coordinate $x_*$ at time $t_*$. If the solution $\psi(x,t)$ is real, the characteristic is also real, leading to the classic result \citep[see e.g.][]{Olver2013-po} of a shock appearing at the earliest possible time $t_*=\min_\xi\left\{-1/\psi_0'(\xi)\right\}$, from which it follows that $\psi_0''(\xi_*)=0$. This highlights that any given (real) initial distribution $\red{\psi_0(x)>0}$ must have a negative slope $\psi_0'<0$ somewhere along the real line for a singularity to develop.  We can also calculate the second derivative of the solution
\begin{equation} \label{eq:dxx_char}
    \pd{\psi}{x}{2}(x,t) = \dfrac{\psi_0''(\xi)}{(1+t\,\psi_0'(\xi))^3}\red{.}
\end{equation}
\red{In the case of real solutions, at the point $x_s(t)\defeq{x_*}-(t_*-t)\psi(x_*,t_*)$ moving with the shock, the second derivative is $\partial_{xx}\psi(x_s(t),t)=\psi_0''(\xi_*)/(1+t\,\psi_0'(\xi_*))^3=0$, which highlights that the profile is locally linear in the moving frame of reference.}

The case of complex solutions is more complicated\red{, since the condition for a singularity to occur must be satisfied for both the real and imaginary parts of a complex-valued $\xi_*$},
\begin{equation} \label{eq:cond_singularity_full}
    1+t_*\re{\psi_0'(\xi_*)}=0, \quad\im{\psi_0'(\xi_*)}=0.
\end{equation}
Previous studies \citep{Thess1996-bc,Crowdy2021-vy,Bickel2022-sa} identified that singularities develop at points where surfactant not only reaches a minimum, but \emph{also} \red{has} a value of zero $\Gamma(x_*,t_*)=0$. Building upon this observation, we limit our analysis to singularities where $\im{\psi(x_*,t_*)}=0$, which in turn leads to a real characteristic $\xi_*$. In that case, we have that $\red{\psi_0'(\xi_*)=u_{s0}'(\xi_*)+\ii\Gamma_0'(\xi_*)}$ and the conditions \eqref{eq:cond_singularity_full} for a singularity to occur are simplified, leading to
\begin{equation} \label{eq:cond_singularity_simplified}
    1+t_*\,u_{s0}'(\xi_*)=0, \quad\Gamma_0'(\xi_*)=0.
\end{equation}
The closure time $t_*$ can then be calculated as follows:
\begin{enumerate}
    \item If the surfactant distribution is sufficiently smooth and zero at a single point $\Gamma_0(x_0)=0$, then such point $x_0$ must be a minimum, and so conditions \eqref{eq:cond_singularity_simplified} then lead to $x_0=\xi_*$ and a singularity time given by  
    \begin{equation} \label{eq:t_star_one_pt}
        t_* = -\dfrac{1}{u_{s0}'(\xi_*)}\text{, with }\xi_*\text{ such that }\Gamma_0'(\xi_*)=\Gamma_0(\xi_*)=0.
    \end{equation}
    While $\Gamma_0'(\xi_*)=0$, the second derivative could be either $\Gamma_0''(\xi_*)>0$ in the case of a quadratic minimum or $\Gamma_0''(\xi_*)=0$ for flatter distributions like, for instance, one with a quartic minimum. For that reason, \eqref{eq:t_star_one_pt} applies for `dimples', described in section \ref{sec:filling_sols_dimple}, and also for some `holes', such as the quartic hole described in section \ref{sec:filling_sols_hole}. In general, once $t_*$ and $\xi_*$ are calculated using \eqref{eq:t_star_one_pt}, one can retrieve the velocity of the singularity $u_{*}=u_{s0}(\xi_*)$ and then the actual position $x_*$ of the singularity using $x_* = \xi_*+t_*u_*$. \red{In the case of a symmetric surfactant distribution} (as in figures \ref{fig:dimple} and \ref{fig:hole_quartic}), the odd interfacial velocity imposes $u_*=0$ and thus the singular point is static with $x_*=\xi_*$. 
    \item If the initial surfactant is zero on a finite interval, then $\Gamma_0'(\xi)=\Gamma_0''(\xi)=0$ on any point of the interval as well.  This is the case for some `holes' such as the rectangular hole and the asymmetric hole from section \ref{sec:filling_sols_hole}. Such distributions lead to singularities at multiple points, since the solution develops a moving front that converges inwards as in figures \ref{fig:hole_rectangular} and \ref{fig:hole_asymmetric}. However, the hole closure time $t_*$ will be determined by the \emph{last} instant in which a singularity occurs, so it can in this case be calculated as
    \begin{equation} \label{eq:t_star_interval}
        t_* = \max_\xi\left\{-\dfrac{1}{u_{s0}'(\xi)}\right\}\text{, and }\xi_*= \underset{\xi}{\arg\max}\left\{-\dfrac{1}{u_{s0}'(\xi)}\right\},
    \end{equation}
    which also implies that $u_{s0}''(\xi_*)=0$. Like in the previous case, the velocity and position of the singularity can in general be retrieved as $u_{*}=u_{s0}(\xi_*)$ and $x_* = \xi_*+t_*u_*$, respectively, and for symmetric distributions (as in figure \ref{fig:hole_rectangular}) we have that $u_*=0$ and $x_*=\xi_*$.
\end{enumerate}

\section{Dictionary of initial conditions} \label{sec:apdx_hilbert_transforms}
\begin{table}
\begin{center}
\def~{\hphantom{0}}
\begin{tabular}{c c c c c c}
     Name & $\Gamma_0(x)$ & $u_{s0}(x)=\hil{\Gamma_0(x)}$ & $\psi_0(z)$ & $M_0$ & $t_*$ \\ \hline
     Cauchy pulse & $\dfrac{1}{1+x^2}$ & $\dfrac{x}{1+x^2}$ & $\red{\dfrac{1}{z-\ii}}$ & $\pi$ & N/A \\[15pt]
     Rectangular pulse & $H(1-|x|)$ & $\dfrac{1}{\pi}\ln\left|\dfrac{x+1}{x-1}\right|$ & $\red{\dfrac{1}{\pi}\,\Ln\hspace{-1pt}\left(\dfrac{z+1}{z-1}\right)}$ & $2$ & N/A \\[15pt]
     Half Cauchy pulse & $\dfrac{H(x)}{1+x^2}$ & $\dfrac{\pi{x}+2\ln|x|}{2\pi(1+x^2)}$ & $\red{\dfrac{\pi(z+\ii)+2\,\Ln(z)}{2\pi(1+z^2)}}$ & $\dfrac{\pi}{2}$ & N/A\\[15pt]
     Quartic pulse & $\dfrac{1}{1+x^4}$ & $\dfrac{x(1+x^2)}{\sqrt{2}\,(1+x^4)}$ & $\red{\dfrac{z-\ii\sqrt{2}}{\sqrt{2}\,(z^2-\ii\sqrt{2}\,z-1)}}$ & $\dfrac{\sqrt{2}\,\pi}{2}$ & N/A\\[15pt]
     Cauchy dimple & $1-\dfrac{1}{1+x^2}$ & $-\dfrac{x}{1+x^2}$ & $\red{\dfrac{\ii z}{z-\ii}}$ & $\infty$ & 1\\[15pt]
     Squared Cauchy dimple & $1-\dfrac{1}{(1+x^2)^2}$ & $-\dfrac{x(3+x^2)}{2(1+x^2)^2}$ & $\red{\dfrac{\ii z(2z-3\ii)}{2(z-\ii)^2}}$ & $\infty$ & $\dfrac{2}{3}$\\[15pt]
     Arctangent dimple & $1-\dfrac{\arctan{x}}{x}$ & $-\dfrac{\ln(1+x^2)}{2x}$ & $\red{\ii-\dfrac{\Ln(1+\ii{z})}{z}}$ & $\infty$ & 2\\[15pt]
     Rectangular hole & $H(|x|-1)$ & $\dfrac{1}{\pi}\ln\left|\dfrac{x-1}{x+1}\right|$ & $\red{\dfrac{1}{\pi}\,\Ln\hspace{-1pt}\left(\dfrac{1-z}{1+z}\right)}$ & $\infty$ & $\dfrac{\pi}{2}$\\[15pt]
     Quartic hole & $1-\dfrac{1}{1+x^4}$ & $-\dfrac{x(1+x^2)}{\sqrt{2}\,(1+x^4)}$ & $\red{\dfrac{z(1+\ii\sqrt{2}{z})}{\sqrt{2}\,(z^2-\ii\sqrt{2}\,z-1)}}$ & $\infty$ & $\sqrt{2}$\\[15pt]
\end{tabular}
\caption{Initial conditions $\Gamma_0(x)$ and $u_{s0}(x)$ used in the article. The fourth column lists the \red{lower}-analytic complex function $\psi_0(z)$, with $z=x+\ii y$, that results in $\red{\psi_0(x)=u_{s0}(x)+\ii\Gamma_0(x)}$ on the real axis $y=0$. The fifth column denotes the mass of surfactant, as defined in equation \eqref{eq:def_M0}, for the case of pulses. The last column specifies the singularity time $t_*$ for holes or dimples, as defined in \eqref{eq:t_star_one_pt} and \eqref{eq:t_star_interval}. $H(x)$ denotes the Heaviside step function.}
\label{tab:hilbert_transforms}
\end{center}
\end{table}
Table \ref{tab:hilbert_transforms} compiles the functional form of profiles $\red{\Gamma_0(x)=\im{\psi_0(x)}}$ and their Hilbert transforms $u_{s0}(x)=\re{\psi_0(x)}=\hil{\Gamma_0(x)}$ used in sections \ref{sec:spreading_sols} and \ref{sec:filling_sols}. In addition, the \red{lower}-analytic complex function $\red{\psi_0(x+\ii{y})}$ that reduces to $\red{\psi_0(x)=u_{s0}(x)+\ii\Gamma_0(x)}$ on the real line ($y=0$) is also provided. This function is required to compute exact solutions to \eqref{eq:Hopf} via the method of characteristics since the solution $\psi(x,t)=\psi_0(x-t\psi(x,t))$ involves evaluations of $\psi_0(z)$ at complex departure points $\red{z}$.

For spreading solutions, multiple pulses can be generated via linear combination \red{$\psi_0(z) = \sum_{n=1}^{N}a_n\,\psi_{0,n}\left((z-c_n)/b_n\right)$ of $N$ shifted and rescaled solutions $\psi_{0,n}$.} The properties of the Hilbert transform \citep[][]{King2009-hr} lead to simple expressions for the total mass $M_0 = \sum_{n=1}^{N}a_nb_nM_{0,n}$ and first moment $M_1 = \sum_{n=1}^{N}a_nb_nc_nM_{0,n}$, where $M_{0,n}$ is the mass of the $n$-th pulse. For the double ($N=2$) quartic pulse in figure \ref{fig:spreading_no_diff_double}, we \red{fix} $a_1=a_2=K$, $b_1=1/3$, $b_2=2/3$, $c_1=-1/2$, $c_2=1$, with $K$ such that $\red{\max[\Gamma_0(x)]=1}$.

Dimple and hole profiles can be readily generated from a pulse \red{$\psi_0^P(z)$ by defining $\psi_0^{H}(z)=\ii-\psi_0^{P}(z)$ for the dimple or hole. More complicated}
functional forms of $\red{\psi_0(z)}$ can be produced in a similar fashion. For instance, the asymmetric hole of figure \ref{fig:hole_asymmetric}, which has an expression that is too long to include in table \ref{tab:hilbert_transforms}, can be built using superposition \red{and shifts} of simpler profiles. If we label the half Cauchy pulse of surfactant as $\psi_{0,A}(z)$, and the rectangular pulse as $\psi_{0,B}(z)$ (both in table \ref{tab:hilbert_transforms}), then the asymmetric hole can be generated as $\red{\psi_0(z)=\ii-\psi_{0,A}(z-1)-\psi_{0,B}(z)}$.

\bibliographystyle{jfm}
\bibliography{paperpile_viscous_Marangoni_spreading}

\begin{thebibliography}{72}
\expandafter\ifx\csname natexlab\endcsname\relax\def\natexlab#1{#1}\fi
\def\au#1{#1} \def\ed#1{#1} \def\yr#1{#1}\def\at#1{#1}\def\jt#1{\textit{#1}} \def\bt#1{#1}\def\bvol#1{\textbf{#1}} \def\vol#1{#1} \def\pg#1{#1} \def\publ#1{#1}\def\arxiv#1{#1}\def\org#1{#1}\def\st#1{\textit{#1}}

\bibitem[Ahmad \& Hansen(1972)]{Ahmad1972-ay}
{\sc \au{Ahmad, J.} \& \au{Hansen, R.S.}} \yr{1972}  \at{{A simple quantitative treatment of the spreading of monolayers on thin liquid films}}.  \jt{J. Colloid Interface Sci.}  \bvol{38}~(3),  \pg{601--604}.

\bibitem[Alpers \& H{\"u}hnerfuss(1989)]{Alpers1989-aj}
{\sc \au{Alpers, W.} \& \au{H{\"u}hnerfuss, H.}} \yr{1989}  \at{{The damping of ocean waves by surface films: A new look at an old problem}}.  \jt{J. Geophys. Res.}  \bvol{94}~(C5),  \pg{6251--6265}.

\bibitem[Baker {\em et~al.\/}(1996)Baker, Li \& Morlet]{Baker1996-gx}
{\sc \au{Baker, G.R.}, \au{Li, X.} \& \au{Morlet, A.C.}} \yr{1996}  \at{{Analytic structure of two 1D-transport equations with nonlocal fluxes}}.  \jt{Physica D}  \bvol{91}~(4),  \pg{349--375}.

\bibitem[Barenblatt(1996)]{Barenblatt1996-hd}
{\sc \au{Barenblatt, G.I.}} \yr{1996} {\em {Scaling, self-similarity, and intermediate asymptotics}\/}.  \publ{Cambridge University Press}.

\bibitem[Bickel \& Detcheverry(2022)]{Bickel2022-sa}
{\sc \au{Bickel, T.} \& \au{Detcheverry, F.}} \yr{2022}  \at{{Exact solutions for viscous Marangoni spreading}}.  \jt{Phys. Rev. E}  \bvol{106}~(4),  \pg{045107}.

\bibitem[Borgas \& Grotberg(1988)]{Borgas1988-hd}
{\sc \au{Borgas, M.S.} \& \au{Grotberg, J.B.}} \yr{1988}  \at{{Monolayer flow on a thin film}}.  \jt{J. Fluid Mech.}  \bvol{193},  \pg{151--170}.

\bibitem[Botte \& Mansutti(2005)]{Botte2005-go}
{\sc \au{Botte, V.} \& \au{Mansutti, D.}} \yr{2005}  \at{{Numerical modelling of the Marangoni effects induced by plankton-generated surfactants}}.  \jt{J. Mar. Syst.}  \bvol{57}~(1),  \pg{55--69}.

\bibitem[Brenner \& Bertozzi(1993)]{Brenner1993-qc}
{\sc \au{Brenner, M.} \& \au{Bertozzi, A.}} \yr{1993}  \at{{Spreading of droplets on a solid surface}}.  \jt{Phys. Rev. Lett.}  \bvol{71}~(4),  \pg{593--596}.

\bibitem[Brenner {\em et~al.\/}(1996)Brenner, Lister \& Stone]{Brenner1996-po}
{\sc \au{Brenner, M.P.}, \au{Lister, J.R.} \& \au{Stone, H.A.}} \yr{1996}  \at{{Pinching threads, singularities and the number 0.0304}}.  \jt{Phys. Fluids}  \bvol{8}~(11),  \pg{2827--2836}.

\bibitem[Breward \& Howell(2002)]{Breward2002-si}
{\sc \au{Breward, C.J.W.} \& \au{Howell, P.D.}} \yr{2002}  \at{{The drainage of a foam lamella}}.  \jt{J. Fluid Mech.}  \bvol{458},  \pg{379--406}.

\bibitem[Cantat {\em et~al.\/}(2013)Cantat, Cohen-Addad, Elias, Graner, H{\"o}hler, Pitois, Rouyer \& Saint-Jalmes]{Cantat2013-se}
{\sc \au{Cantat, I.}, \au{Cohen-Addad, S.}, \au{Elias, F.}, \au{Graner, F.}, \au{H{\"o}hler, R.}, \au{Pitois, O.}, \au{Rouyer, F.} \& \au{Saint-Jalmes, A.}} \yr{2013} {\em {Foams: Structure and Dynamics}\/}.  \publ{Oxford University Press}.

\bibitem[Chae {\em et~al.\/}(2005)Chae, C{\'o}rdoba, C{\'o}rdoba \& Fontelos]{Chae2005-st}
{\sc \au{Chae, D.}, \au{C{\'o}rdoba, A.}, \au{C{\'o}rdoba, D.} \& \au{Fontelos, M.A.}} \yr{2005}  \at{{Finite time singularities in a 1D model of the quasi-geostrophic equation}}.  \jt{Adv. Math.}  \bvol{194}~(1),  \pg{203--223}.

\bibitem[Cox(2012)]{Cox2012-kk}
{\sc \au{Cox, D.A.}} \yr{2012} {\em {Galois Theory}\/}.  \publ{John Wiley \& Sons}.

\bibitem[Crowdy(2021{\natexlab{{\em a\/}}})]{Crowdy2021-ef}
{\sc \au{Crowdy, D.G.}} \yr{2021{\natexlab{{\em a\/}}}}  \at{{Exact solutions for the formation of stagnant caps of insoluble surfactant on a planar free surface}}.  \jt{J. Eng. Math.}  \bvol{131}~(1),  \pg{10}.

\bibitem[Crowdy(2021{\natexlab{{\em b\/}}})]{Crowdy2021-vy}
{\sc \au{Crowdy, D.G.}} \yr{2021{\natexlab{{\em b\/}}}}  \at{{Viscous Marangoni flow driven by insoluble surfactant and the complex Burgers equation}}.  \jt{SIAM J. Appl. Math.}  \bvol{81}~(6),  \pg{2526--2546}.

\bibitem[Crowdy {\em et~al.\/}(2023)Crowdy, Curran \& Papageorgiou]{Crowdy2023-aj}
{\sc \au{Crowdy, D.G.}, \au{Curran, A.E.} \& \au{Papageorgiou, D.T.}} \yr{2023}  \at{{Fast reaction of soluble surfactant can remobilize a stagnant cap}}.  \jt{J. Fluid Mech.}  \bvol{969},  \pg{A8}.

\bibitem[Cuenot {\em et~al.\/}(1997)Cuenot, Magnaudet \& Spennato]{Cuenot1997-mh}
{\sc \au{Cuenot, B.}, \au{Magnaudet, J.} \& \au{Spennato, B.}} \yr{1997}  \at{{The effects of slightly soluble surfactants on the flow around a spherical bubble}}.  \jt{J. Fluid Mech.}  \bvol{339},  \pg{25--53}.

\bibitem[Day {\em et~al.\/}(1998)Day, Hinch \& Lister]{Day1998-tw}
{\sc \au{Day, R.F.}, \au{Hinch, E.J.} \& \au{Lister, J.R.}} \yr{1998}  \at{{Self-similar capillary pinchoff of an inviscid fluid}}.  \jt{Phys. Rev. Lett.}  \bvol{80}~(4),  \pg{704--707}.

\bibitem[Eggers(1993)]{Eggers1993-iu}
{\sc \au{Eggers, J.}} \yr{1993}  \at{{Universal pinching of 3D axisymmetric free-surface flow}}.  \jt{Phys. Rev. Lett.}  \bvol{71}~(21),  \pg{3458--3460}.

\bibitem[Eggers(2000)]{Eggers2000-ep}
{\sc \au{Eggers, J.}} \yr{2000}  \at{{Singularities in droplet pinching with vanishing viscosity}}.  \jt{SIAM J. Appl. Math.}  \bvol{60}~(6),  \pg{1997--2008}.

\bibitem[Eggers \& Fontelos(2008)]{Eggers2008-lk}
{\sc \au{Eggers, J.} \& \au{Fontelos, M.A.}} \yr{2008}  \at{{The role of self-similarity in singularities of partial differential equations}}.  \jt{Nonlinearity}  \bvol{22}~(1),  \pg{R1}.

\bibitem[Eggers \& Fontelos(2015)]{Eggers2015-pz}
{\sc \au{Eggers, J.} \& \au{Fontelos, M.A.}} \yr{2015} {\em {Singularities: Formation, Structure and Propagation}\/}.  \publ{Cambridge University Press}.

\bibitem[Eggers \& Fontelos(2020)]{Eggers2020-wo}
{\sc \au{Eggers, J.} \& \au{Fontelos, M.A.}} \yr{2020}  \at{{Selection of singular solutions in non-local transport equations}}.  \jt{Nonlinearity}  \bvol{33}~(1),  \pg{325}.

\bibitem[Erinin {\em et~al.\/}(2023)Erinin, Liu, Liu, Mostert, Deike \& Duncan]{Erinin2023-rb}
{\sc \au{Erinin, M.A.}, \au{Liu, C.}, \au{Liu, X.}, \au{Mostert, W.}, \au{Deike, L.} \& \au{Duncan, J.H.}} \yr{2023}  \at{{The effects of surfactants on plunging breakers}}.  \jt{J. Fluid Mech.}  \bvol{972},  \pg{R5}.

\bibitem[Frumkin \& Levich(1947)]{Frumkin1947-fm}
{\sc \au{Frumkin, A.N.} \& \au{Levich, V.G.}} \yr{1947}  \at{{On surfactants and interfacial motion}}.  \jt{Zh. Fiz. Khim.}  \bvol{21},  \pg{1183--1204}.

\bibitem[Gaver \& Grotberg(1990)]{Gaver1990-wj}
{\sc \au{Gaver, D.P.} \& \au{Grotberg, J.B.}} \yr{1990}  \at{{The dynamics of a localized surfactant on a thin film}}.  \jt{J. Fluid Mech.}  \bvol{213},  \pg{127--148}.

\bibitem[Gaver \& Grotberg(1992)]{Gaver1992-xu}
{\sc \au{Gaver, D.P.} \& \au{Grotberg, J.B.}} \yr{1992}  \at{{Droplet spreading on a thin viscous film}}.  \jt{J. Fluid Mech.}  \bvol{235},  \pg{399--414}.

\bibitem[de~Gennes {\em et~al.\/}(2004)de~Gennes, Brochard-Wyart \& Qu{\'e}r{\'e}]{De_Gennes2004-hx}
{\sc \au{de~Gennes, P.-G.}, \au{Brochard-Wyart, F.} \& \au{Qu{\'e}r{\'e}, D.}} \yr{2004} {\em {Capillarity and Wetting Phenomena}\/}.  \publ{Springer New York}.

\bibitem[Giga \& Kohn(1985)]{Giga1985-ht}
{\sc \au{Giga, Y.} \& \au{Kohn, R.V.}} \yr{1985}  \at{{Asymptotically self‐similar blow‐up of semilinear heat equations}}.  \jt{Commun. Pure Appl. Math.}  \bvol{38}~(3),  \pg{297--319}.

\bibitem[Giga \& Kohn(1987)]{Giga1987-xg}
{\sc \au{Giga, Y.} \& \au{Kohn, R.V.}} \yr{1987}  \at{{Characterizing blowup using similarity variables}}.  \jt{Indiana Univ. Math. J.}  \bvol{36}~(1),  \pg{1--40}.

\bibitem[Gratton \& Minotti(1990)]{Gratton1990-uc}
{\sc \au{Gratton, J.} \& \au{Minotti, F.}} \yr{1990}  \at{{Self-similar viscous gravity currents: phase-plane formalism}}.  \jt{J. Fluid Mech.}  \bvol{210},  \pg{155--182}.

\bibitem[Griffith(1962)]{Griffith1962-oh}
{\sc \au{Griffith, R.M.}} \yr{1962}  \at{{The effect of surfactants on the terminal velocity of drops and bubbles}}.  \jt{Chem. Eng. Sci.}  \bvol{17}~(12),  \pg{1057--1070}.

\bibitem[Grotberg {\em et~al.\/}(1995)Grotberg, Halpern \& Jensen]{Grotberg1995-ph}
{\sc \au{Grotberg, J.B.}, \au{Halpern, D.} \& \au{Jensen, O.E.}} \yr{1995}  \at{{Interaction of exogenous and endogenous surfactant: spreading-rate effects}}.  \jt{J. Appl. Physiol.}  \bvol{78}~(2),  \pg{750--756}.

\bibitem[de~la Hoz \& Fontelos(2008)]{De_la_Hoz2008-hp}
{\sc \au{de~la Hoz, F.} \& \au{Fontelos, M.A.}} \yr{2008}  \at{{The structure of singularities in nonlocal transport equations}}.  \jt{J. Phys. A: Math. Theor.}  \bvol{41}~(18),  \pg{185204}.

\bibitem[Huppert(1982)]{Huppert1982-nj}
{\sc \au{Huppert, H.E.}} \yr{1982}  \at{{The propagation of two-dimensional and axisymmetric viscous gravity currents over a rigid horizontal surface}}.  \jt{J. Fluid Mech.}  \bvol{121},  \pg{43--58}.

\bibitem[Jensen(1994)]{Jensen1994-hf}
{\sc \au{Jensen, O.E.}} \yr{1994}  \at{{Self‐similar, surfactant‐driven flows}}.  \jt{Phys. Fluids}  \bvol{6}~(3),  \pg{1084--1094}.

\bibitem[Jensen(1995)]{Jensen1995-mu}
{\sc \au{Jensen, O.E.}} \yr{1995}  \at{{The spreading of insoluble surfactant at the free surface of a deep fluid layer}}.  \jt{J. Fluid Mech.}  \bvol{293},  \pg{349--378}.

\bibitem[Jensen \& Grotberg(1992)]{Jensen1992-eq}
{\sc \au{Jensen, O.E.} \& \au{Grotberg, J.B.}} \yr{1992}  \at{{Insoluble surfactant spreading on a thin viscous film: shock evolution and film rupture}}.  \jt{J. Fluid Mech.}  \bvol{240},  \pg{259--288}.

\bibitem[Jensen \& Grotberg(1993)]{Jensen1993-xa}
{\sc \au{Jensen, O.E.} \& \au{Grotberg, J.B.}} \yr{1993}  \at{{The spreading of heat or soluble surfactant along a thin liquid film}}.  \jt{Physics of Fluids A: Fluid Dynamics}  \bvol{5}~(1),  \pg{58--68}.

\bibitem[Kaneelil {\em et~al.\/}(2022)Kaneelil, Pahlavan, Xue \& Stone]{Kaneelil2022-kq}
{\sc \au{Kaneelil, P.R.}, \au{Pahlavan, A.A.}, \au{Xue, N.} \& \au{Stone, H.A.}} \yr{2022}  \at{{Three-dimensional self-similarity of coalescing viscous drops in the thin-film regime}}.  \jt{Phys. Rev. Lett.}  \bvol{129}~(14),  \pg{144501}.

\bibitem[King(2009{\natexlab{{\em a\/}}})]{King2009-hr}
{\sc \au{King, F.W.}} \yr{2009{\natexlab{{\em a\/}}}} {\em {Hilbert Transforms: Volume 1}\/}.  \publ{Cambridge University Press}.

\bibitem[King(2009{\natexlab{{\em b\/}}})]{King2009-bq}
{\sc \au{King, F.W.}} \yr{2009{\natexlab{{\em b\/}}}} {\em {Hilbert Transforms: Volume 2}\/}.  \publ{Cambridge University Press}.

\bibitem[Leal(2007)]{Leal2007-zv}
{\sc \au{Leal, L.G.}} \yr{2007} {\em {Advanced Transport Phenomena: Fluid Mechanics and Convective Transport Processes}\/}.  \publ{Cambridge University Press}.

\bibitem[Lister \& Kerr(1989)]{Lister1989-so}
{\sc \au{Lister, J.R.} \& \au{Kerr, R.C.}} \yr{1989}  \at{{The propagation of two-dimensional and axisymmetric viscous gravity currents at a fluid interface}}.  \jt{J. Fluid Mech.}  \bvol{203},  \pg{215--249}.

\bibitem[Liu \& Duncan(2003)]{Liu2003-vf}
{\sc \au{Liu, X.} \& \au{Duncan, J.H.}} \yr{2003}  \at{{The effects of surfactants on spilling breaking waves}}.  \jt{Nature}  \bvol{421}~(6922),  \pg{520--523}.

\bibitem[Lucassen \& Van Den~Tempel(1972)]{Lucassen1972-nw}
{\sc \au{Lucassen, J.} \& \au{Van Den~Tempel, M.}} \yr{1972}  \at{{Dynamic measurements of dilational properties of a liquid interface}}.  \jt{Chem. Eng. Sci.}  \bvol{27}~(6),  \pg{1283--1291}.

\bibitem[Manikantan \& Squires(2020)]{Manikantan2020-lh}
{\sc \au{Manikantan, H.} \& \au{Squires, T.M.}} \yr{2020}  \at{{Surfactant dynamics: hidden variables controlling fluid flows}}.  \jt{J. Fluid Mech.}  \bvol{892}.

\bibitem[Matar \& Craster(2009)]{Matar2009-jt}
{\sc \au{Matar, O.K.} \& \au{Craster, R.V.}} \yr{2009}  \at{{Dynamics of surfactant-assisted spreading}}.  \jt{Soft Matter}  \bvol{5}~(20),  \pg{3801--3809}.

\bibitem[Morlet(1998)]{Morlet1998-rn}
{\sc \au{Morlet, A.C.}} \yr{1998}  \at{{Further properties of a continuum of model equations with globally defined flux}}.  \jt{J. Math. Anal. Appl.}  \bvol{221}~(1),  \pg{132--160}.

\bibitem[Olver {\em et~al.\/}(2010)Olver, Ozier, Boisvert \& Clark]{Olver2010-qx}
{\sc \au{Olver, F.W.J.}, \au{Ozier, D.W.}, \au{Boisvert, R.F.} \& \au{Clark, C.W.}}, ed. \yr{2010} {\em {NIST Handbook of Mathematical Functions}\/}.  \publ{Cambridge University Press}.

\bibitem[Olver(2014)]{Olver2013-po}
{\sc \au{Olver, P.J.}} \yr{2014} {\em {Introduction to Partial Differential Equations}\/}.  \publ{Springer}.

\bibitem[Palaparthi {\em et~al.\/}(2006)Palaparthi, Papageorgiou \& Maldarelli]{Palaparthi2006-ud}
{\sc \au{Palaparthi, R.}, \au{Papageorgiou, D.T.} \& \au{Maldarelli, C.}} \yr{2006}  \at{{Theory and experiments on the stagnant cap regime in the motion of spherical surfactant-laden bubbles}}.  \jt{J. Fluid Mech.}  \bvol{559},  \pg{1--44}.

\bibitem[Park(1991)]{Park1991-ja}
{\sc \au{Park, C.-W.}} \yr{1991}  \at{{Effects of insoluble surfactants on dip coating}}.  \jt{J. Colloid Interface Sci.}  \bvol{146}~(2),  \pg{382--394}.

\bibitem[Peaudecerf {\em et~al.\/}(2017)Peaudecerf, Landel, Goldstein \& Luzzatto-Fegiz]{Peaudecerf2017-dz}
{\sc \au{Peaudecerf, F.J.}, \au{Landel, J.R.}, \au{Goldstein, R.E.} \& \au{Luzzatto-Fegiz, P.}} \yr{2017}  \at{{Traces of surfactants can severely limit the drag reduction of superhydrophobic surfaces}}.  \jt{Proc. Natl. Acad. Sci. U.S.A.}  \bvol{114}~(28),  \pg{7254--7259}.

\bibitem[Pozrikidis(1992)]{Pozrikidis1992-gr}
{\sc \au{Pozrikidis, C.}} \yr{1992} {\em {Boundary Integral and Singularity Methods for Linearized Viscous Flow}\/}.  \publ{Cambridge University Press}.

\bibitem[Qu{\'e}r{\'e}(1999)]{Quere1999-bv}
{\sc \au{Qu{\'e}r{\'e}, D.}} \yr{1999}  \at{{Fluid coating on a fiber}}.  \jt{Annu. Rev. Fluid Mech.}  \bvol{31}~(1),  \pg{347--384}.

\bibitem[Sadhal \& Johnson(1983)]{Sadhal1983-fg}
{\sc \au{Sadhal, S.S.} \& \au{Johnson, R.E.}} \yr{1983}  \at{{Stokes flow past bubbles and drops partially coated with thin films. Part 1. Stagnant cap of surfactant film -- exact solution}}.  \jt{J. Fluid Mech.}  \bvol{126},  \pg{237--250}.

\bibitem[Schechter \& Farley(1963)]{Schechter1963-ne}
{\sc \au{Schechter, R.S.} \& \au{Farley, R.W.}} \yr{1963}  \at{{Interfacial tension gradients and droplet behavior}}.  \jt{Can. J. Chem. Eng.}  \bvol{41}~(3),  \pg{103--107}.

\bibitem[Scriven \& Sternling(1960)]{Scriven1960-ie}
{\sc \au{Scriven, L.E.} \& \au{Sternling, C.V.}} \yr{1960}  \at{{The Marangoni effects}}.  \jt{Nature}  \bvol{187}~(4733),  \pg{186--188}.

\bibitem[Slim \& Huppert(2004)]{Slim2004-fo}
{\sc \au{Slim, A.C.} \& \au{Huppert, H.E.}} \yr{2004}  \at{{Self-similar solutions of the axisymmetric shallow-water equations governing converging inviscid gravity currents}}.  \jt{J. Fluid Mech.}  \bvol{506},  \pg{331--355}.

\bibitem[Song {\em et~al.\/}(2018)Song, Song, Hu, Du, Du, Choi \& Rothstein]{Song2018-hi}
{\sc \au{Song, D.}, \au{Song, B.}, \au{Hu, H.}, \au{Du, X.}, \au{Du, P.}, \au{Choi, C.-H.} \& \au{Rothstein, J.P.}} \yr{2018}  \at{{Effect of a surface tension gradient on the slip flow along a superhydrophobic air-water interface}}.  \jt{Phys. Rev. Fluids}  \bvol{3}~(3),  \pg{033303}.

\bibitem[Strogatz(2018)]{Strogatz2018-tt}
{\sc \au{Strogatz, S.H.}} \yr{2018} {\em {Nonlinear Dynamics and Chaos: With Applications to Physics, Biology, Chemistry, and Engineering}\/}.  \publ{CRC Press}.

\bibitem[Temprano-Coleto {\em et~al.\/}(2023)Temprano-Coleto, Smith, Peaudecerf, Landel, Gibou \& Luzzatto-Fegiz]{Temprano-Coleto2023-hb}
{\sc \au{Temprano-Coleto, F.}, \au{Smith, S.M.}, \au{Peaudecerf, F.J.}, \au{Landel, J.R.}, \au{Gibou, F.} \& \au{Luzzatto-Fegiz, P.}} \yr{2023}  \at{{A single parameter can predict surfactant impairment of superhydrophobic drag reduction}}.  \jt{Proc. Natl. Acad. Sci. U.S.A.}  \bvol{120}~(3),  \pg{e2211092120}.

\bibitem[Thess(1996)]{Thess1996-bc}
{\sc \au{Thess, A.}} \yr{1996}  \at{{Stokes flow at infinite Marangoni number: exact solutions for the spreading and collapse of a surfactant}}.  \jt{Phys. Scr.}  \bvol{1996}~(T67),  \pg{96}.

\bibitem[Thess {\em et~al.\/}(1995)Thess, Spirn \& J{\"u}ttner]{Thess1995-pm}
{\sc \au{Thess, A.}, \au{Spirn, D.} \& \au{J{\"u}ttner, B.}} \yr{1995}  \at{{Viscous flow at infinite Marangoni number}}.  \jt{Phys. Rev. Lett.}  \bvol{75}~(25),  \pg{4614--4617}.

\bibitem[Thess {\em et~al.\/}(1997)Thess, Spirn \& J{\"u}ttner]{Thess1997-qi}
{\sc \au{Thess, A.}, \au{Spirn, D.} \& \au{J{\"u}ttner, B.}} \yr{1997}  \at{{A two-dimensional model for slow convection at infinite Marangoni number}}.  \jt{J. Fluid Mech.}  \bvol{331},  \pg{283--312}.

\bibitem[Trinschek {\em et~al.\/}(2018)Trinschek, John \& Thiele]{Trinschek2018-ev}
{\sc \au{Trinschek, S.}, \au{John, K.} \& \au{Thiele, U.}} \yr{2018}  \at{{Modelling of surfactant-driven front instabilities in spreading bacterial colonies}}.  \jt{Soft Matter}  \bvol{14}~(22),  \pg{4464--4476}.

\bibitem[Wang {\em et~al.\/}(1999)Wang, Papageorgiou \& Maldarelli]{Wang1999-kk}
{\sc \au{Wang, Y.}, \au{Papageorgiou, D.T.} \& \au{Maldarelli, C.}} \yr{1999}  \at{{Increased mobility of a surfactant-retarded bubble at high bulk concentrations}}.  \jt{J. Fluid Mech.}  \bvol{390},  \pg{251--270}.

\bibitem[Wasserman \& Slattery(1969)]{Wasserman1969-cg}
{\sc \au{Wasserman, M.L.} \& \au{Slattery, J.C.}} \yr{1969}  \at{{Creeping flow past a fluid globule when a trace of surfactant is present}}.  \jt{AIChE J.}  \bvol{15}~(4),  \pg{533--547}.

\bibitem[Wu {\em et~al.\/}(2024)Wu, Duprat \& Stone]{Wu2024-ew}
{\sc \au{Wu, K.}, \au{Duprat, C.} \& \au{Stone, H.A.}} \yr{2024}  \at{{Capillary rise in sharp corners: not quite universal}}.  \jt{J. Fluid Mech.}  \bvol{978},  \pg{A26}.

\bibitem[Zheng {\em et~al.\/}(2018)Zheng, Fontelos, Shin \& Stone]{Zheng2018-bu}
{\sc \au{Zheng, Z.}, \au{Fontelos, M.A.}, \au{Shin, S.} \& \au{Stone, H.A.}} \yr{2018}  \at{{Universality in the nonlinear leveling of capillary films}}.  \jt{Phys. Rev. Fluids}  \bvol{3}~(3),  \pg{032001}.

\bibitem[Zhong {\em et~al.\/}(2019)Zhong, Ketelaar, Braun, Begley \& King-Smith]{Zhong2019-kd}
{\sc \au{Zhong, L.}, \au{Ketelaar, C.F.}, \au{Braun, R.J.}, \au{Begley, C.G.} \& \au{King-Smith, P.E.}} \yr{2019}  \at{{Mathematical modelling of glob-driven tear film breakup}}.  \jt{Math. Med. Biol.}  \bvol{36}~(1),  \pg{55--91}.

\end{thebibliography}

\end{document}